\documentclass[aps,prd,nofootinbib,superscriptaddress,notitlepage,preprintnumbers]{revtex4-1}

\usepackage{graphicx}
\usepackage{amsmath,graphicx,verbatim,epsfig}
\usepackage{epstopdf}
\usepackage[utf8x]{inputenc}
\usepackage[colorlinks=true,linkcolor=blue,citecolor=blue]{hyperref}
\usepackage[usenames,dvipsnames]{color}
\usepackage{bm}

\newcommand{\la}{\lambda}

\newcommand{\nn}{\nonumber}

\newcommand{\GeV}{\rm GeV}

\newcommand{\be}{\begin{equation}}
\newcommand{\ee}{\end{equation}}
\newcommand{\bea}{\begin{eqnarray}}
\newcommand{\eea}{\end{eqnarray}}
\newcommand{\balign}{\begin{align}}
\newcommand{\ealign}{\end{align}}
\newcommand{\as}{\alpha_s}

\newcommand{\bg}{\begin{gather}}
\newcommand{\foma}{\end{gather}}

\newcommand{\noopsort}[1]{}

\newcommand{\vecb}[1]{\mbox{\boldmath $#1$}}
\newcommand{\vecbe}[1]{\mbox{\boldmath ${\scriptstyle #1}$}}

\def\L{\Lambda}

\def\z{\zeta}

\def\<{\langle}
\def\>{\rangle}

\def\a{\alpha}
\def\b{\beta}
\def\g{\gamma}  \def\G{\Gamma}
\def\d{\delta}  
\def\l{\lambda}   \def\L{\Lambda}
\def\s{\sigma}

\def\m{\mu}

\def\z{\zeta}

\def\({\left(}
\def\[{\left[}
\def\){\right)}
\def\]{\right]}

\def\cos{\hbox{cos}}
\def\sin{\hbox{sin}}
\def\ln{\hbox{ln}}

\def\Qslash{Q\!\!\!\!\slash}

\def \le { \left    }
\def \ri { \right }

\def\lqcd{\L_{\rm QCD}}

\begin{document}

\title{Non-perturbative QCD effects in $q_T$ spectra of Drell-Yan and $Z$-boson production }

\author{Umberto D'Alesio}
\email{umberto.dalesio@ca.infn.it}
\affiliation{Dipartimento di Fisica, Universit\`a di Cagliari,
             I-09042 Monserrato (CA), Italy}
\affiliation{INFN, Sezione di Cagliari,
             C.P. 170, I-09042 Monserrato (CA), Italy}

\author{Miguel G. Echevarria}
\email{m.g.echevarria@nikhef.nl}
\affiliation{Nikhef and Department of Physics and Astronomy,
VU University Amsterdam,
De Boelelaan 1081, NL-1081 HV Amsterdam, the Netherlands}

\author{Stefano Melis}
\email{stefano.melis@to.infn.it}
\affiliation{Dipartimento di Fisica, Universit\`a di Torino,
Via P. Giuria 1, I-10125 Torino, Italy}

\author{Ignazio Scimemi}
\email{ignazios@fis.ucm.es}
\affiliation{Departamento de F\'isica Te\'orica II,
Universidad Complutense de Madrid (UCM),
28040 Madrid, Spain}

\date{\today}


\begin{abstract}
The factorization theorems for transverse momentum distributions of dilepton/boson production, recently formulated  by Collins and Echevarria-Idilbi-Scimemi in terms of well-defined transverse momentum dependent distributions (TMDs), allows for a systematic and quantitative analysis of non-perturbative QCD effects of the cross sections involving these quantities.
In this paper we perform a global fit using all current available data for Drell-Yan and $Z$-boson production at hadron colliders within this framework.
The perturbative calculable pieces of our estimates are included using a complete resummation at next-to-next-to-leading-logarithmic accuracy.
Performing the matching of transverse momentum distributions onto the standard collinear parton distribution functions and recalling that the corresponding matching coefficient can be partially exponentiated, we find that this exponentiated part is spin-independent and resummable.
We argue that the inclusion of higher order perturbative pieces is necessary when data from lower energy scales are analyzed.
We consider non-perturbative corrections both to the intrinsic nucleon structure and to the evolution kernel and find that the non-perturbative part of the TMDs could be parametrized in terms of a minimal set of parameters (namely 2-3).
When all corrections are included the global fit so performed gives a $\chi^2/{\rm d.o.f.} \lesssim 1$ and a very precise prediction for vector boson production at the Large Hadron Collider (LHC).
\end{abstract}

\preprint{NIKHEF 2014-023}

\maketitle

\section{Introduction}

The Drell-Yan (DY) and vector  boson ($Z$) production are fundamental processes for the Large Hadron Collider (LHC) physics as they represent a  basic test for strong and electroweak interactions. In this paper we focus on the transverse momentum distributions of the leptons coming from Drell-Yan and $Z$-boson production when the sum of the transverse momentum of the dilepton pair is small compared to the center of mass energy and/or the invariant mass of the neutral boson. This particular regime is sensitive to non-perturbative QCD effects and has received a particular attention for a long time~\cite{Collins:1984kg,Landry:2002ix,Bozzi:2010xn,Becher:2011xn}.
The new developments of the theory of transverse momentum dependent distributions (shortly referred to as TMDs)~\cite{Collins:2011zzd,GarciaEchevarria:2011rb,Echevarria:2012js} forces us to a re-thinking of the analysis of non-perturbative QCD effects.
In fact, the proper definition of the TMDs allows a systematic analysis of non-perturbative QCD in the observed transverse momentum distributions. These effects are fundamental to fix the precision of current and future experiments, including those at the LHC and at an Electron Ion Collider (EIC).
The cross sections that we consider in this work can be formulated schematically according to the factorization formula~\cite{Collins:2011zzd,GarciaEchevarria:2011rb,Echevarria:2012js}
\begin{align}
\frac{d\s}{dq_T} \sim
H(Q^2,\m^2)
\int d^2\vecb k_{AT}\, d^2\vecb k_{BT} \,
F_A(x_A,\vecb k_{AT};\z_A,\m)\,
F_B(x_B,\vecb k_{BT};\z_B,\m)\,
\d^{(2)}(\vecb k_{AT}+\vecb k_{BT}-\vecb q_{T})\,,
\end{align}
where $F_{A,B}$ are the transverse momentum dependent parton distribution functions (TMDPDFs).
They depend on the dilepton invariant mass trough the scales $\z_A$ and $\z_B$\footnote{In Ref.~\cite{Echevarria:2012js,Echevarria:2014rua} the authors used the equivalent notation $\z_A=Q^2/\alpha$ and $\z_B=Q^2\alpha$, where $\alpha$ is the soft function splitting parameter.}, being $\z_A\z_B=Q^4$, the intrinsic parton transverse momenta, the factorization scale $\m$ and the Bjorken momentum fractions.
Finally, $H$ is the hard factor, which is spin independent and can be calculated adopting the standard perturbation theory.
Notice that this formula can be easily extended to the case of semi-inclusive deep inelastic scattering (SIDIS) by  replacing the two TMDPDFs by one TMDPDF and one TMD fragmentation function (TMDFF); analogously for electron-positron annihilation we would have the convolution of two TMDFFs~\footnote{The hard factor, $H$, should also be changed accordingly, depending on the time-like or space-like nature of the momentum of the relevant boson.}.

The TMDPDFs, like the ordinary collinear or integrated parton distribution functions (PDFs), are quantities which can be evolved between two different scales, so that all the non-perturbative QCD information encoded in a TMDPDF at one scale can be directly used at a different one by means of the appropriate evolutor.
In impact parameter space we have
\begin{align}
\label{eq:FRF}
\tilde F(x,b_T;\z_f,\m_f)=\tilde R(b_T;\z_i,\m_i,\z_f,\m_f) \tilde F(x,b_T;\z_i,\m_i)
\,,
\end{align}
where $\tilde R$ is the evolution kernel.
Notice that here below we use $\m_{i,f}^2=\z_{i,f}=Q_{i,f}^2$ for simplicity.

In order to make the inverse Fourier transform of the TMDPDF in momentum space one has to define some prescriptions to treat the non-perturbative regime in impact parameter space. To fix the ideas we can refer to some aspects of the well-known Collins-Soper-Sterman (CSS) resummation framework~\cite{Collins:1984kg}.
In this approach the initial scale for the evolution is chosen directly in impact parameter space, namely $Q_i=C/b_T$ where $C$ is a numerical constant.
This choice induces a Landau pole singularity, which is cured introducing a cutoff in impact parameter space, the well-known $b^*$-prescription.
Thus, the Landau pole is avoided by freezing the strong coupling constant in impact parameter space.
However, this procedure, which is an important aspect of this approach, modifies the TMDs also in the region where their perturbative expression is supposed to hold.

In this paper we take a different path, fixing the initial scale in Eq.~(\ref{eq:FRF}) directly as $Q_i\sim Q_0+q_T$, where $Q_0$ is a minimal energy scale at the border between perturbative and non-perturbative regimes
($Q_0\sim 2 \;{\rm GeV}$).
We find that the effect of the Landau pole is actually minimized by this choice.
This confirms the result of Ref.~\cite{Echevarria:2012pw}, where it was argued that the completely resummed evolution kernel could be implemented in a large region of the impact parameter space.
We also study the impact of a $Q$-dependent non-perturbative part using a specific model.
We anticipate here that this correction plays a role only in the fit at next-to-next-to-leading-logarithmic (NNLL) accuracy.

In Ref.~\cite{Echevarria:2012pw} the authors have deduced also the anomalous dimension of the TMDs up to next-to-next-to-leading order (NNLO), which is at the same level as the usual analysis of perturbative QCD in terms of collinear parton distribution functions.
An expression of the evolutor suitable for higher order analysis is also provided in the same paper.
This information is used now in this work to perform a global fit of data at NNLL, which, to our knowledge, represents a novelty in the field.

The functional form of the TMDPDF, viable both for low-energy Drell-Yan processes (say, processes with invariant masses $Q\sim$ a few GeV) and for high energies (say $Q\sim M_Z$ or bigger) is still under debate.
In this work we discuss and show how to construct a functional form of the TMDPDF which incorporates the full resummations with the highest available perturbative ingredients (for the moment a complete N$^3$LL analysis is not possible as one misses, for instance, the four-loop cusp anomalous dimension).

A fundamental piece of information for TMDs comes from their limit when the transverse momentum is much larger than the hadronization scale ($q_T\gg \Lambda_{\rm QCD}$).
In this tail, the TMDPDF matches onto a Wilson coefficient and a PDF and this fact, to the best of our knowledge, is used in all models.
It is well known from the literature that this matching coefficient of the TMDPDF onto the PDF can be partially exponentiated (see~\cite{Kodaira:1981nh} and, more recently, \cite{Becher:2011xn}).
Such exponentiated part is spin independent (i.e. the same for all leading-twist TMDs) and obeys a differential equation very similar to the one found for the evolution kernel.
We provide an analytic solution of this equation in such a way that the exponentiated part of the matching coefficient is resummed consistently with the evolution kernel.

We find that this splitting of a TMDPDF into a Wilson coefficient and a PDF needs important non-perturbative corrections that we parametrize through a model, with a minimal set of parameters (namely two, called $\l_{1,2}$ in the text).
These corrections are independent of the dilepton invariant mass $Q$.
We have performed a study of several model forms and found that the (corrected) exponential form outlined in Section~\ref{theory} is the best suited for the minimization of the $\chi^2$ of our global fit.
On top of this we have also explored the case of a $Q$-dependent non-perturbative contribution, which requires the use of a third parameter ($\l_3$ in the following).
When this correction is included we find a general improvement of the fit, especially for the low-energy data.
For this reason we think that this extra piece could play a relevant role also when studying the data from fixed-target SIDIS experiments, that are mostly collected at low $Q^2$ values.
Notice that this kind of correction is universal, in the sense that it is the same for all TMDs.

The data that we have studied in this work come from the following experiments:
E288~ \cite{Ito:1980ev},
E605~\cite{Moreno:1990sf},
R209~\cite{Antreasyan:1981uv},
CDF Run I~\cite{Affolder:1999jh},
D0 Run I~\cite{Abbott:1999yd,Abbott:1999wk},
CDF Run II~\cite{Aaltonen:2012fi}, D0 Run II~\cite{Abazov:2007ac}. These data cover a large interval in center of mass energies, dilepton invariant masses and rapidities.  The specific features (and role) of each experiment are further described below in the text.

The paper is organized as follows: in Section~\ref{theory} we present in some detail the theoretical approach,  discussing the construction of the TMDPDFs and the main aspects of the cross sections that we are going to study. We also comment on some features of the data under consideration.
In Section~\ref{pheno} we show our phenomenological analysis with a selection of the most interesting results, in Section~\ref{sec:lit} we comment of  previous works on the subject and then, in Section~\ref{concl}, we draw our conclusions.

\section{Theoretical Framework}
\label{theory}

As stated above, the processes which are relevant in this work are DY and $Z$-boson production, $A\,B\rightarrow \g^*\rightarrow \ell^+\ell^-$, $A\,B\rightarrow Z\rightarrow \ell^+\ell^-$, where $A$ and $B$ are the incoming hadrons whose center of mass energy is denoted as $\sqrt{s}$ in the following.
In the case of DY, say $\g^*$ production, one can write the differential cross section as
\begin{align}
\frac{d\sigma}{dQ^2\, dq_T^2\,dy} &=
\sum_q \sigma_q^\g H(Q^2,\m^2)
\int \frac{d^2\vecb b_T}{4 \pi} e^{-i\vecbe q_T \cdot \vecbe b_T}
\tilde F_{q/A}(x_A,b_T;\z_A,\m)\,
\tilde F_{\bar q/B}(x_B,b_T;\z_B,\m)
+Y_\g\,,
\end{align}
where, adopting the standard notation,
\begin{align}
\sigma_q^\g &=\frac{4\pi \alpha^2_{\rm em}}{3 N_c Q^2 s} e_q^2\,,\quad
x_{A,B}=\sqrt{\tau_\g}\, e^{\pm y}\,,\quad
\tau_\g =\frac{Q^2+q_T^2}{s}\simeq \frac{Q^2}{s}\,.
\end{align}
For the case of $Z$-boson production the analogous observable is
\begin{align}
\frac{d\sigma}{dq_T^2\,dy} &=
\sum_q \sigma_q^Z H(M_Z^2,\m^2)
\int \frac{d^2\vecb b_T}{4\pi} e^{-i\vecbe q_T\cdot \vecbe b_T}
\tilde F_{q/A}(x_A,b_T;\z_A,\m)\,
\tilde F_{\bar q/B}(x_B,b_T;\z_B,\m) + Y_Z\,,
\end{align}
where
\begin{align}
\sigma_q^Z &=
\frac{4\pi^2\alpha_{\rm em}}{ N_c  s} \frac{(1-2|e_q|\sin^2\theta_W)^2
+4 e_q^2\sin^4\theta_W}{8\,\sin^2\theta_W\cos^2\theta_W}
B_R(Z\rightarrow \ell^+\ell^-)\,,\quad
x_{A,B}=\sqrt{\tau_Z}\, e^{\pm y}\,,\quad
\tau_Z =\frac{M_Z^2+q_T^2}{s}\simeq \frac{M_Z^2}{s}\,.
\end{align}
The approximate value of $\tau_{\gamma, Z}$ holds only when $Q^2\gg q_T^2$, which is a good approximation for all sets of data we have used.
The term $Y_{\g,Z}$ is the usual remnant piece used to recover the complete perturbative limit of the cross section at high transverse momenta.
 In Ref.~\cite{Becher:2011xn}  it was  shown that the $Y_{\g,Z}$ terms provide extremely small corrections, so that, accordingly, we have neglected their contribution in the present work.
For the hard coefficient we use $H(Q^2,\m^2)=|C_V(Q^2,\m^2)|^2$, where $C_V$  is the Wilson coefficient that can be extracted from the finite terms of the calculation of the (full QCD) quark form factor in pure dimensional regularization and is known up to three loops~\cite{Baikov:2009bg,Gehrmann:2010ue}.
At one-loop we have:
\begin{align}
\label{eq:hardpart}
C_V(Q^2,\m^2) &=
1 + \frac{\as}{4\pi}\le[
-\ln^2\frac{-Q^2}{\m^2} +3 \ln\frac{-Q^2}{\m^2} - 8 + \frac{\pi^2}{6}
\ri]
\,.
\end{align}

Since, as shown in the equation above, the hard coefficient has an explicit dependence on $-Q^2$, in this work we have used the $\pi$-resummed coefficient as suggested in Ref.~\cite{Ahrens:2008qu}.
This resummation ensures a slightly better convergence of the perturbative expansion in the hard coefficient, although the effects are not numerically relevant as for the Higgs boson production.
In the following the factorization scale $\m$ is fixed at the same value as the invariant mass of the exchanged vector boson, $Q$ or $M_Z$, so that the TMDPDFs are completely evolved up to this energy scale and the logarithms in the hard part are minimized.
For some sets of data one has to integrate on the rapidity $y$ within the range $|y|\leq y_{max}^i=-\frac{1}{2}\ln\tau_i$, $i=\g,Z$.

In order to finally build the TMDPDFs we need to include the completely resummed evolution kernel and discuss the operator product expansion of the TMDPDFs onto the PDFs. This is done in the following sections.

\subsection{Evolution kernel}
\label{sec:EK}

The basic quantities which encode all QCD non-perturbative information are the evolved TMDPDFs.
The evolution of a generic quark-TMDPDF is given by Eq.~(\ref{eq:FRF}), where the evolution kernel $\tilde R$ is~\cite{Echevarria:2012pw}
\begin{align}\label{eq:tmdkernel}
\tilde R(b_T;\z_i,\m_i,\z_f,\m_f) &=
\exp\le\{
\int_{\m_i}^{\m_f} \frac{d\bar\m}{\bar\m}
\g_F\le(\as(\bar\m),\ln\frac{\z_f}{\bar\m^2} \ri)
\ri\}
\le( \frac{\z_f}{\z_i} \ri)^{-D\le(b_T;\m_i\ri)}
\,.
\end{align}
In this equation $\g_F$ is the anomalous dimension of the TMDPDF and is related to the anomalous dimension of the hard factor as explained in Ref.~\cite{GarciaEchevarria:2011rb}.
The function $D$ can be resummed to obtain the $D^R$ function, that will replace $D$ in Eq.~(\ref{eq:tmdkernel}), via the renormalization group equation
\begin{align}
\label{eq:Deq}
\frac{d D}{d\ln\mu}&=\G_{\rm cusp}\,,
\end{align}
where $\G_{\rm cusp}$ is the cusp anomalous dimension.
Solving this equation one finds
\begin{align}
\label{eq:DR}
D(b_T;\m)&=
D(b_T;\m_b) +
\int_{\a_s(\m_b)}^{\a_s(\mu)}d\a^\prime
\frac{\G_{\rm cusp}(\a^\prime)}{\b(\a^\prime)}
\end{align}
and re-writing $\a_s(\m_b)$ as a series in $\a_s(\mu)$, where  $\m_b=2 e^{-\g_E}/b_T$ and $\b(\as)$ is the QCD $\b$-function~\cite{Echevarria:2012pw}, one finds
\begin{align}\label{eq:resummedD}
 D^R(b_T;\m) &=
-\frac{\Gamma_0}{2\beta_0}\ln(1-X)
+ \frac{1}{2}\le(\frac{a_s}{1-X}\ri) \le[
- \frac{\beta_1\Gamma_0}{\beta_0^2} (X+\ln(1-X))
+\frac{\Gamma_1}{\beta_0} X\ri]
\nn\\
&+ \frac{1}{2}
\le(\frac{a_s}{1-X}\ri)^2\le[
2d_2(0)
+\frac{\Gamma_2}{2\beta_0}(X (2-X))
+\frac{\beta_1\Gamma_1}{2 \beta_0^2} \le( X (X-2)-2 \ln (1-X)\ri)
+\frac{\beta_2\Gamma_0}{2\beta_0^2} X^2 \ri.\nn\\
&+ \le. \frac{\beta_1^2\Gamma_0}{2\beta_0^3} (\ln^2(1-X)-X^2)\right]\,,
\end{align}
where we have used the notation
\begin{align}
a_s &=\frac{\as(\m)}{4\pi} \,,
\quad\quad\quad
X=a_s \b_0 L_T \,,
\quad\quad\quad
L_T=\ln\frac{\m^2 b_T^2}{4e^{-2\g_E}} = \ln \frac{\m^2}{\m^2_b}
\,.
\end{align}

The final form of the full resummed kernel is
\begin{align}\label{eq:tmdkernelf}
\tilde R^{\rm pert}(b_T;\z_i,\m_i,\z_f,\m_f) &=
\exp\le\{
\int_{\m_i}^{\m_f} \frac{d\bar\m}{\bar\m}
\g_F\le(\as(\bar\m),\ln\frac{\z_f}{\bar\m^2} \ri)\ri\}
\le( \frac{\z_f}{\z_i} \ri)^
{-D^R\le(b_T;\m_i\ri)
}\,.
\end{align}

The resummation of the evolution kernel is valid only up to a certain maximum value of the impact parameter~\cite{Echevarria:2012pw, Echevarria:2014rua}, beyond which the evolution kernel becomes completely non-perturbative.
Given that the evolution kernel can be derived from the soft matrix element~\cite{Echevarria:2014rua}, which is process and spin independent, the non-perturbative correction to the evolution kernel is also process and spin independent. This correction to the $D^R$ term provides then a correction to the evolution kernel  depending on $\ln (\z_f/\z_i)$, as can be deduced from Eq.~(\ref{eq:tmdkernel}).
In this work we have tried to fix this non-perturbative part phenomenologically, once the complete perturbative resummation is performed.
The complete form of the evolution kernel is then
\begin{align}\label{eq:tmdkernelt}
\tilde R(b_T;\z_i,\m_i,\z_f,\m_f) =
\tilde R^{\rm pert}(b_T;\z_i,\m_i,\z_f,\m_f)
\le( \frac{\z_f}{\z_i} \ri)^
{-D^{\rm NP}\le(b_T\ri)
}\,,
\end{align}
where $D^{\rm NP}$ is the non-perturbative piece of the $D$ function.

\subsection{Operator Product Expansion of the TMDPDF onto a PDF, spin-independent exponentiation and resummation}
In this section we recall the perturbative limit of the TMDPDFs for high transverse momenta, building the TMDPDF consistently.

The main constraint on the structure of TMDPDFs comes from their limit at high transverse momentum, $q_T\gg \lqcd$.
In this case the TMDPDFs split into a Wilson coefficient and a PDF, see the specific literature for its proof~\cite{Collins:2011zzd,GarciaEchevarria:2011rb}.
Taking into account this perturbative tail we start writing the TMDPDF for a generic nucleon $N$ as
\begin{align}
\label{eq:coeff1}
\tilde F_{q/N}(x,b_T;\z,\m)&=
\left(\frac{\z b_T^2}{4 e^{-2\g_E} }\right)^{-D(b_T;\m)}
\sum_j
\int_{x}^{1}\frac{dz}{z}
\tilde{C}^{\Qslash}_{q\leftarrow j}(x/z,b_T;\mu)\,
f_{j/N}(z;\m)
+{\cal O}(b_T\lqcd)
\,.
\end{align}

The coefficients $\tilde C^{\Qslash}$ for the quark-quark and the quark-gluon channel at one loop are~\cite{GarciaEchevarria:2011rb}
\begin{align}
\tilde C^{\Qslash}_{q\leftarrow q'}(z,b_T;\m) &=
\delta(1-z)\d_{qq'}
+2 a_s C_F \left[1-z-\delta(1-z)\left(-\frac{1}{2} L_T^2-\frac{3}{2} L_T+\frac{\pi^2}{12}
\right)-{\cal P}_{q\leftarrow q'} L_T\right] \d_{qq'}\,,\\
\tilde{C}^{\Qslash}_{q\leftarrow g}(z,b_T;\m) &=
2a_s T_F\le[ 2z(1-z) - {\cal P}_{q\leftarrow g} L_T \right]
\,,
\end{align}
and the one-loop splitting kernels appearing above are given by
\begin{align}
{\cal P}_{q\leftarrow q}&=
\left(\frac{1+z^2}{1-z}\right)_+
= \frac{2z}{(1-z)_+} + (1-z) + \frac{3}{2}\d(1-z)
\,,
\nn\\
{\cal P}_{q\leftarrow g}&=
z^2 + (1-z)^2
\,.
\end{align}

The important point is that under renormalization group equation one has
\begin{align}
\frac{d}{d\ln \mu} \tilde{C}^{\Qslash}_{q\leftarrow j}(z,b_T;\m) &=
(\G_{\rm cusp}L_T - \g^V)
\tilde{C}^{\Qslash}_{q\leftarrow j}(z,b_T;\m)
-\sum_i\int_z^1\frac{d\xi}{\xi}
\tilde{C}^{\Qslash}_{q\leftarrow i}(\xi,b_T;\m)\, {\cal P}_{i\leftarrow j}(z/\xi)\,,
\end{align}
where the double logarithms can be partially exponentiated in a way similar to what presented in Ref.~\cite{Kodaira:1981nh} (see also Ref.~\cite{Becher:2010tm}):
\begin{align}
\tilde{C}^{\Qslash}_{q\leftarrow j}(z,b_T;\m) \equiv
\exp\le[h_\G(b_T;\m)-h_\g(b_T;\m)\ri]\hat{C}_{q\leftarrow j}(z,b_T;\m)\,,
\end{align}
where
\begin{align} \label{eq:h}
\frac{d h_\G}{d\ln \m}&=\G_{\rm cusp}L_T\;;\quad\quad
\frac{d h_\g}{d\ln \m}=\g^V\, .
\end{align}
The coefficients of the perturbative expansions of $\G_{\rm cusp}$ and $\g^V$ can be found in the Appendix~C of Ref.~\cite{Echevarria:2012pw}.

Choosing $h_{\G,\g}(b_T;\m_b)=0$ we have, at fixed order,
\begin{align}
h_{\G,\g}&=\sum_n h_{\G,\g}^{(n)} \left( \frac{\a_s}{4\pi}\right)^n\,,\nn\\
h_\G^{(1)} &=\frac{1}{4} L_T^2 \G_0\,,\quad\quad
h_\G^{(2)} =\frac{1}{12} (L_T^3 \G_0\b_0+3L_T^2 \G_1)\,,\nn\\
h_\G^{(3)}&=\frac{1}{24}(L_T^4 \G_0\b_0^2+2L_T^3 \G_0\b_1+4L_T^3 \G_1\b_0+6 L_T^2 \G_2)\,,
\nn\\
h_\g^{(1)}&=
\frac{\g_0}{2 \b_0} \le(\b_0 L_T \ri)
\,,
\quad \quad
h_\g^{(2)} =\frac{\g_0}{4\b_0} \le(\b_0 L_T \ri)^2
+ \le(\frac{\g_1}{2\b_0} \ri) \le(\b_0 L_T \ri)\,,\nn\\
h_\g^{(3)} &=
\frac{\g_0}{6\b_0} \le(\b_0 L_T \ri)^3
+ \frac{1}{2}\le(\frac{\g_1}{\b_0} + \frac{1}{2}\frac{\g_0\b_1}{\b_0^2}  \ri)
\le(\b_0 L_T \ri)^2+ \frac{1}{2} \le( \frac{\g_2}{\b_0}\ri) \le(\b_0 L_T \ri)
\,,
\end{align}
and correspondingly
\begin{align}
\hat{C}_{q\leftarrow q'}(z,b_T;\m) &=
\delta(1-z)\d_{qq'}
+2 a_s C_F \left[1-z-\delta(1-z)\frac{\pi^2}{12}
-{\cal P}_{q\leftarrow q'} L_T\right]\d_{qq'}
\,,
\nn\\
\hat{C}_{q\leftarrow g}(z,b_T;\m) &=
2a_s T_F\le[ 2z(1-z) - {\cal P}_{q\leftarrow g} L_T \right]\,.
\end{align}

Thus, the partially resummed TMDPDF can be written as
\begin{align}
\label{eq:TMDFO}
\tilde F_{q/N}(x,b_T;\z,\m)
&=
\left(\frac{\z b_T^2}{4 e^{-2\g_E} }\right)^{-D(b_T;\m)}
e^{h_\G(b_T;\m)-h_\g(b_T;\m)}
\sum_j
\int_{x}^{1}\frac{dz}{z}
\hat C_{q\leftarrow j}(x/z,b_T;\m) \,
f_{j/N}(z;\m)
+{\cal O}(b_T\lqcd)
\,.
\end{align}

The expression above still contains large logarithms $L_T$ that need to be resummed when $\a_s L_T $ is of order 1. This resummation is also necessary to have an expression consistent with the fully resummed evolution kernel derived in the previous Section. Solving the evolution equations for $D$, Eq.~(\ref{eq:Deq}), and for $h_\G$ and $h_\g$,
Eq.~(\ref{eq:h}), we find respectively Eq.~(\ref{eq:DR})
and
\begin{align}
\label{eq:hR}
h_\G(b_T;\m) &=
h_\G(b_T;\m_b)
+ \int_{\m_b}^{\m}\frac{d\bar\m}{\bar\m} \G_{\rm cusp}L_T\,,\nn \\
h_\g(b_T;\m) &=
h_\g(b_T;\m_b)
+ \int_{\m_b}^{\m}\frac{d\bar\m}{\bar\m} \g^V\,.
\end{align}
The method of integration is the same used in Ref.~\cite{Echevarria:2012pw} for the evaluation of $D^R$, and it can  be immediately applied to find $h_\g^R$ as the equation for $h_\g$ (second line of Eq.~(\ref{eq:hR})) has the same functional form as Eq.~(\ref{eq:DR}), so that
\begin{align}\label{eq:hg}
 h_\g^R(b_T;\m) &=
-\frac{\g_0}{2\beta_0}\ln(1-X)
+ \frac{1}{2}\le(\frac{a_s}{1-X}\ri) \le[
- \frac{\beta_1\g_0}{\beta_0^2} (X+\ln(1-X))
+\frac{\g_1}{\beta_0} X\ri]
\nn\\
&+ \frac{1}{2}
\le(\frac{a_s}{1-X}\ri)^2\le[
\frac{\g_2}{2\beta_0}(X (2-X))
+\frac{\beta_1\g_1}{2 \beta_0^2} \le( X (X-2)-2 \ln (1-X)\ri)
+\frac{\beta_2\g_0}{2\beta_0^2} X^2 \ri.
\nn\\
&\le.
+\frac{\beta_1^2\g_0}{2\beta_0^3} (\ln^2(1-X)-X^2)\right]\,.
\end{align}
In the case of $h_\G$ we note that $L_T=\int_{\a_s(\m_b)}^{\a_s(\mu)}\frac{d\a}{\b(\a)}$,
and in this way Eq.~(\ref{eq:hR}) (first line) can be solved as
\begin{align}
h_\G^R(b_T;\m)&=\int_{\a_s(\m_b)}^{\a_s(\mu)}d\a^\prime\frac{\Gamma_{\rm cusp}(\a^\prime)}{\beta(\a^\prime)}
\int_{\a_s(\m_b)}^{\a^\prime}\frac{d\a}{\b(\a)}\,.
\end{align}
One can perform the above integration first expanding the $\b$-function and finally re-writing $\a_s(\m_b)$ in terms of $\a_s(\m)$ at the correct order, as shown in Ref.~\cite{Echevarria:2012pw}.
The result is
\begin{align}
h_\Gamma^R(b_T;\m)&=\frac{ \Gamma_0 (X-(X-1) \ln (1-X))}{2 a_s \beta_0^2}+
 \frac{\beta_1 \Gamma_0 \left(2 X+\ln ^2(1-X)+2 \ln (1-X)\right)-2
   \beta_0 \Gamma_1 (X+\ln (1-X))}{4 \beta_0^3}
\nn\\
&
+ \frac{a_s }{4  \beta_0^4 (1-X)}\left(\beta_0^2 \Gamma_2 X^2-\beta_0
   (\beta_1 \Gamma_1 (X (X+2)+2 \ln (1-X))+\beta_2
   \Gamma_0 ((X-2) X+2 (X-1) \ln (1-X)))\right.\nn
\\&\left.
+\beta_1^2 \Gamma_0
   (X+\ln (1-X))^2\right)\,.
\end{align}

Joining all pieces together we get the fully resummed perturbative part of the TMDPDF:
\begin{align}
\tilde F_{q/N}^{\rm pert}(x,b_T;\z,\m)
&=
\left(\frac{\z b_T^2}{4 e^{-2\g_E} }\right)^{-D^R(b_T;\m)}
e^{h_\G^R(b_T;\m)-h_\g^R(b_T;\m)}
\sum_j
\int_{x}^{1}\frac{dz}{z}
\hat C_{q\leftarrow j}(x/z,b_T;\m)\, f_{j/N}(z;\m)\,.
\label{eq:Fpert}
\end{align}

An important point to underline in this formula is that the whole factor in front of the integral is universal among all leading-twist TMDPDFs.
In fact, $D^R$ is a piece of the universal evolution kernel~\cite{Echevarria:2014rua}, and $h_{\G,\g}^R$ are deduced respectively from the cusp and the TMDPDF anomalous dimensions, which are independent of the specific TMDPDF.
This fact shows that the treatment of this factor is not restricted to the unpolarized TMDPDF, but can be applied as well to all (leading-twist) polarized TMDPDFs.
To the best of our knowledge this observation has never been used for the explicit construction of transverse momentum dependent distributions.

\begin{figure}[h]
 \begin{center}
 \includegraphics[width=.45\textwidth, angle=0,natwidth=610,natheight=642]{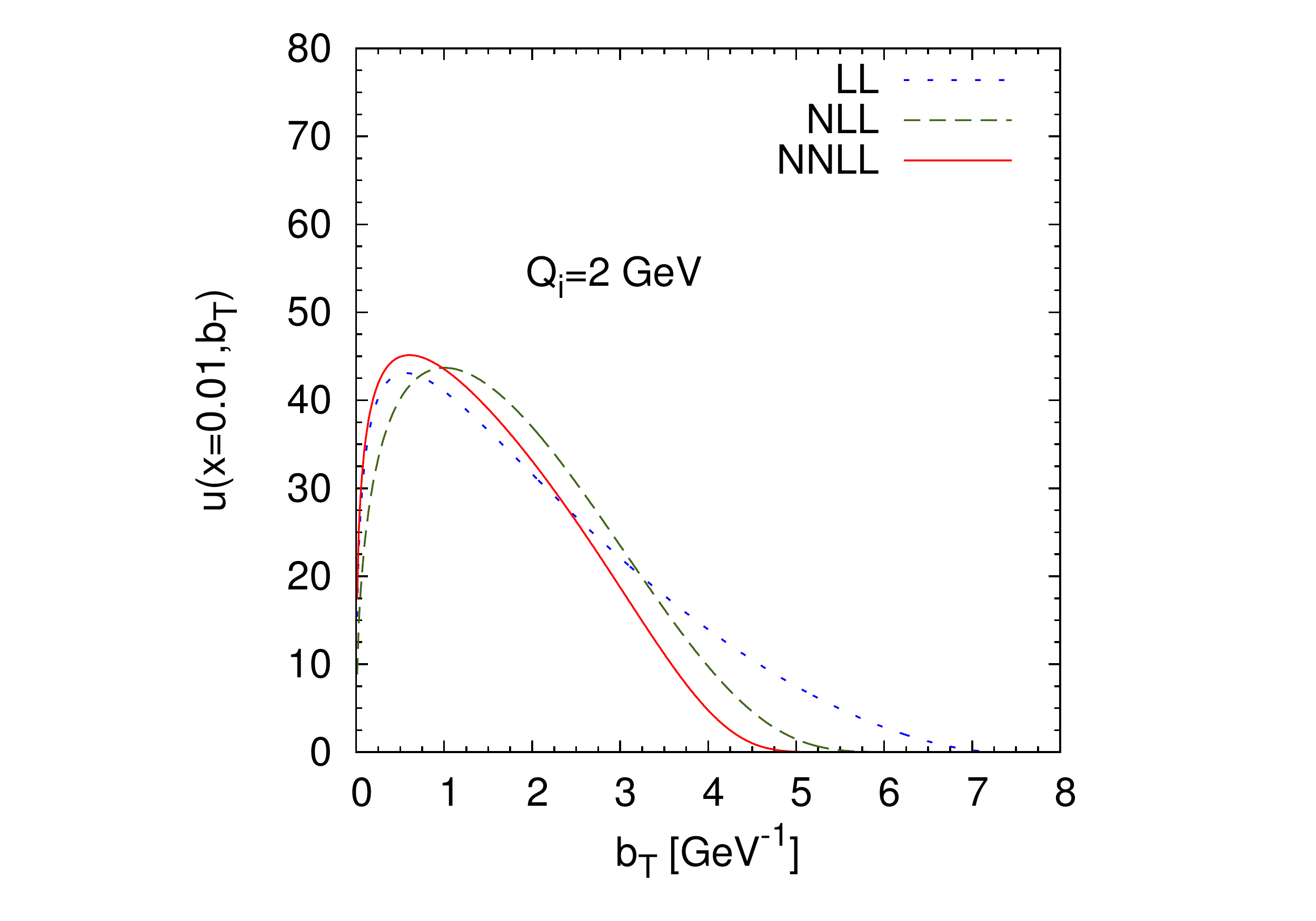}
  \includegraphics[width=.45\textwidth, angle=0,natwidth=610,natheight=642]{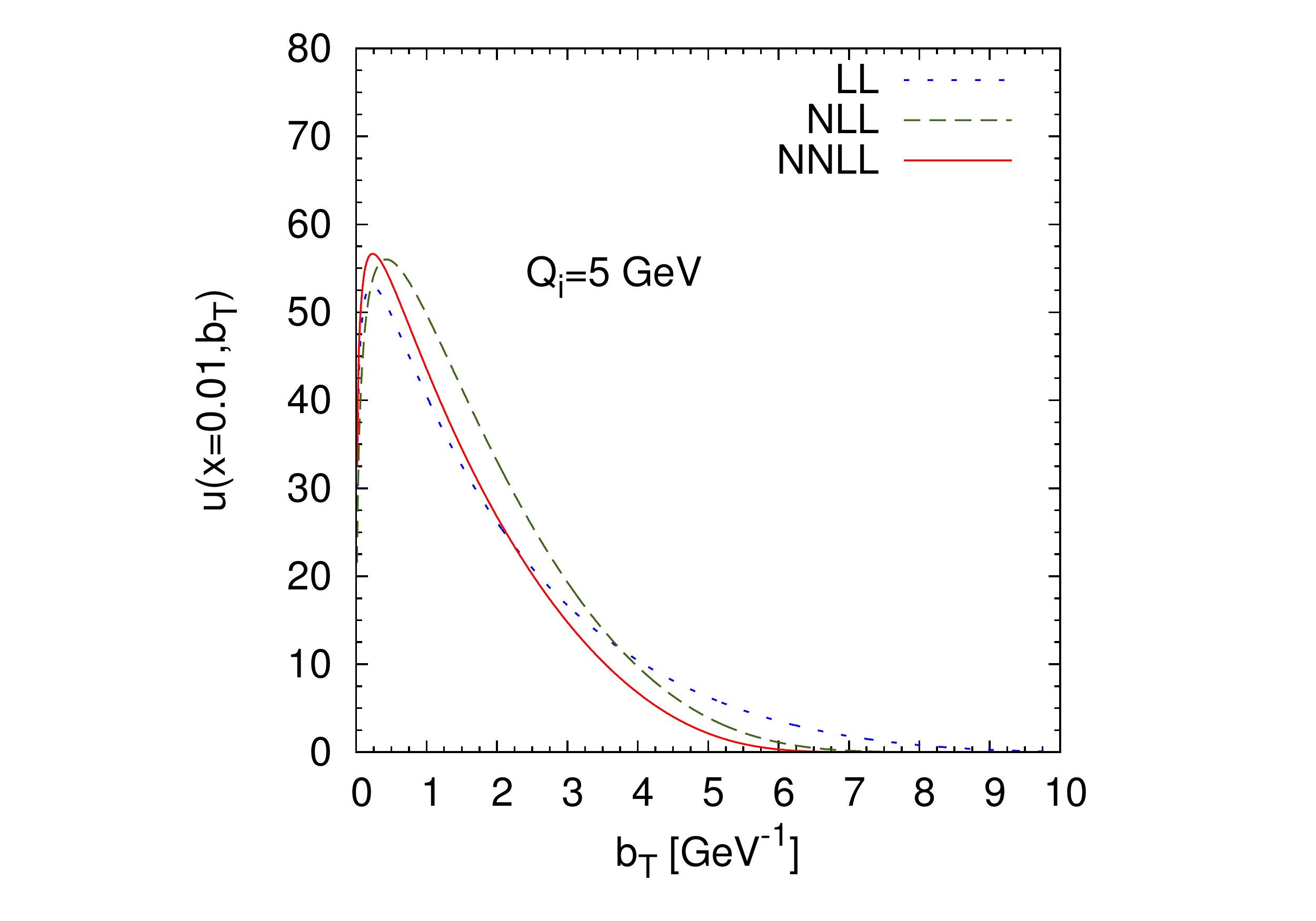}
 \caption{Perturbative part for the up-quark TMDPDF at the initial (fixed) scale $Q_i=2,\; 5$ GeV, $x=0.01$ and various orders in logarithmic resummations as given in Eq.~(\ref{eq:Fpert}).
\label{fig:Fpert}}
 \end{center}
 \end{figure}

An explicit representation of Eq.~(\ref{eq:Fpert}) is given in Fig.~\ref{fig:Fpert} for a fixed scale $\m^2=\z=Q_i^2$.
We notice that even for such a low-energy scale as $Q_i\sim 2$ GeV, the region  of high values of the  impact parameter is suppressed.
A further suppression is found when the non-perturbative input for the TMDPDF is considered, as detailed in the next Section.
We notice that the change between the next-to-leading-log (NLL) and NNLL curves is driven by the matching coefficient $\hat C_{q\leftarrow j}$, whose perturbative order in each curve is detailed in Table~\ref{tab:resummation} (In order to properly follow the power counting one has to keep in mind that we consider large logarithms $L=\ln(Q/q_T)$ so that $\as L\sim 1$).
In this table we report also the orders in $\a_s$ included in each step of the logarithmic resummation.
Comparing the two plots in Fig.~\ref{fig:Fpert} we can also observe that at high values of the impact parameter, where we expect bigger non-perturbative effects, the values of the TMDPDF are basically independent of $Q_i$.
On the other hand for higher values of $Q_i$ and low values of $b_T$ the TMDPDF is particularly enhanced.

\begin{table}[b]
\begin{center}
\begin{tabular}{|c|c|c|c|c|c|c|c|}\hline\hline
Order & $H$ & $\hat C_{q\leftarrow j}$ & $\G_{\rm cusp}$ & $\g^V$ &
$D^R$ & $h_\G^R$ & $h_\g^R$
\\
\hline
LL &$\as^0$&$\as^0$&$\as^1$&$\as^0$&$\as^0$&$\as^{-1}$&$0$
\\
\hline
NLL &$\as^0$&$\as^0$& $\as^2$ & $\as^1$&$\as^1$&$\as^0$&$\as^0$
\\
\hline
NNLL&$\as^1$&$\as^1$&$\as^3$&$\as^2$&$\as^2$&$\as^1$&$\as^1$\\
\hline\hline
\end{tabular}
\caption{Perturbative orders in logarithmic resummations.\label{tab:resummation}}
\end{center}
\end{table}

\subsection{Scale choices and non-perturbative inputs for the TMDPDF}

Despite the resummations performed above the formulas so obtained are still of perturbative origin, and as such they have to be corrected by non-perturbative contributions.
The full resummations allow nevertheless an improvement of the convergence of the perturbative series on a large portion of the impact parameter space.\\
Below we definitely switch to the  notation: $\z_{i,f}=Q_{i,f}^2$ and $\z=Q^2$.

Parametrizing the non-perturbative large-$b_T$ region of the TMDPDF, we write it at some initial scale $Q_i$ as
\begin{align}
\label{eq:FqN1}
\tilde F_{q/N}(x,b_T;Q_i^2,\m_i) =
{\tilde F}^{\rm pert}_{q/N}(x,b_T;Q_i^2,\m_i)\,
{\tilde F}^{\rm NP}_{q/N}(x,b_T;Q_i)
\,,
\end{align}
where ${\tilde F}^{\rm NP}_{q/N}(x,b_T;Q_i)$ is the non-perturbative part of the TMDPDF with
\begin{align}
\label{eq:fgfg}
{\tilde F}^{\rm NP}_{q/N}(x,b_T;Q_i)\equiv
{\tilde F}^{\rm NP}_{q/N}(x,b_T)\left(\frac{Q_i^2}{Q_0^2}\right)^{-D^{\rm NP}(b_T)}
\,.
\end{align}
Notice that in the equation above we have parametrized the non-perturbative contribution in the same way as we did for the evolution kernel in Eq.~\eqref{eq:tmdkernelt}.

We observe that ${\tilde F}^{\rm pert}_{q/N}(x,b_T;Q_i^2,\m_i)$ goes to zero for high values of $b_T$ at fixed $Q_i$, in agreement with our expectations, see Fig.~\ref{fig:Fpert}, and the non-perturbative contributions are not expected to alter this behavior.

Another relevant issue in Eq.~(\ref{eq:FqN1}) concerns the choice of the scales $Q_i$ and $\m_i$.
In order to minimize the value of the logarithms we choose $\m_i=Q_i$.
Next we notice that the splitting into a coefficient and a PDF is valid only at high transverse momentum, so that we expect that the choice $Q_i= Q_0+q_T$ (where $Q_0$ is a fixed low scale) minimizes the logarithms generated by this splitting.
The scale $Q_0$ works as a minimum matching scale between the TMDPDF and the PDF, such that it sits at the border between the perturbative and non-perturbative regimes; in particular we choose $Q_0\sim 2~\GeV$.

As explained below Eq.~\eqref{eq:hardpart}, in the cross section we fix the factorization scale $\m=Q$, so that collecting all results in Eqs.~(\ref{eq:FRF}),~(\ref{eq:tmdkernelt}) and~(\ref{eq:FqN1}) we can write the resummed TMDPDF that enters into the factorization theorem as
\begin{align}
\label{eq:FRF2}
\tilde F_{q/N}(x,b_T;Q^2,Q)&=
\tilde R^{\rm pert}(b_T;(Q_0+q_T)^2,Q_0+q_T,Q^2,Q)\,
{\tilde F}^{\rm pert}_{q/N}(x,b_T;(Q_0+q_T)^2,Q_0+q_T)\,
{\tilde F}^{\rm NP}_{q/N}(x,b_T;Q)
\,.
\end{align}
We point out that in the last factor in Eq.~(\ref{eq:FRF2}), ${\tilde F}^{\rm NP}_{q/N}(x,b_T;Q)$, we have included the non-perturbative $Q$ dependence coming from the evolution kernel (Eq.~(\ref{eq:tmdkernelt})),
\begin{align}
\label{eq:FNP4}
{\tilde F}^{\rm NP}_{q/N}(x,b_T;Q)=
{\tilde F}^{\rm NP}_{q/N}(x,b_T;Q_i) \left(\frac{Q^2}{Q_i^2}\right)^{-D^{\rm NP}(b_T)}\ .
\end{align}
More explicitly, the TMDPDF is implemented as
\begin{align}
\label{eq:Ffull}
\tilde F_{q/N}(x,b_T;Q^2,Q)
&=
\exp\le\{
\int_{Q_i}^{Q} \frac{d\bar\m}{\bar\m}
\g_F\le(\as(\bar\m),\ln\frac{Q^2}{\bar\m^2} \ri)\ri\}
\left(\frac{Q^2 b_T^2}{4 e^{-2\g_E} }\right)^{-D^R(b_T;Q_i)}
\nn\\
&\times
e^{h_\G^R(b_T;Q_i)-h_\g^R(b_T;Q_i)}
\sum_j
\int_{x}^{1}\frac{dz}{z}
\hat C_{q\leftarrow j}(x/z,b_T;Q_i)\, f_{j/N}(z;Q_i)\,
{\tilde F}^{\rm NP}_{q/N}(x,b_T;Q)
\,,
\end{align}
where, as we already mentioned, $Q_i=Q_0+q_T$.

In order to fix the arguments of the non-perturbative part ${\tilde F}^{\rm NP}$, we need to consider the following constraints:
\begin{itemize}
\item It must correct the behavior of ${\tilde F}^{\rm pert}_{q/N}$ at large values of $b_T$, where the perturbative expansion looses its convergence properties and the Landau pole singularity shows up, both in the evolution kernel and in the matching coefficient of the TMDPDF onto the PDF.
\item It has to be such that
\begin{align}
\label{eq:blimit}
\lim_{b_T\rightarrow 0}{\tilde F}^{\rm NP}_{q/N}=1\,,
\end{align}
in order to guarantee that the perturbative series is not altered where its convergence properties are sound.
\end{itemize}
We have not included a dependence on $x$, as data eventually do not need such correction and to keep the model simple enough.
In Eq.~(\ref{eq:blimit}) we are assuming that the values of $x$ are not extremely small (say $x>10^{-3}$), in which case the whole TMD formalism should be re-considered.

We have studied several parametrizations of the non-perturbative part (Gaussian, polynomial, etc.) and the final one which better provides a good fit of the data, with the minimum set of parameters and $D^{\rm NP}=0$, is
\begin{align}
\label{eq:FqN3}
{\tilde F}^{\rm NP}_{q/N}(x,b_T;Q) &=
e^{-\l_1 b_T}\le(1+\l_2 b_T^2\ri)
\,.
\end{align}
As discussed below in the text the data for $Z$-boson production are basically sensitive just to the parameter $\l_1$, that is to the exponential factor and not to the second power-like term that, controlling the large-$b_T$ region, is more sensitive to small-$q_T$ data.
The global fit so performed allows to fix, to a certain precision, the value of this non-perturbative constant. In other words, this fit can be used to fix the amount of non-perturbative QCD corrections in the transverse momentum spectra.
As commented above, the parameter $\l_2$ corrects the behavior of the TMDPDF at high values of $b_T$ and results necessary to describe the data at low dilepton invariant mass and low $q_T$.

Considering now a nonzero $D^{\rm NP}$, this results in a $Q$-dependent factor in the non-perturbative model (see the studies of Refs.~\cite{Korchemsky:1994is,Tafat:2001in} and more recently Refs.~\cite{Collins:2011zzd,Echevarria:2014rua}).
Thus, from Eqs.~(\ref{eq:FNP4}) and (\ref{eq:FqN3}), by setting $D^{\rm NP}=\l_3 b_T^2/2$, we have
\begin{align}
\label{eq:FqN3m}
{\tilde F}^{\rm NP}_{q/N}(x,b_T;Q)
&=
e^{-\l_1 b_T}\le(1+\l_2 b_T^2\ri)\le(\frac{Q^2}{Q_0^2}\ri)^{-\frac{\l_3}{2} b_T^2}\,.
\end{align}

\begin{figure}[h]
 \begin{center}
 \includegraphics[width=.6\textwidth, angle=0,natwidth=610,natheight=642]{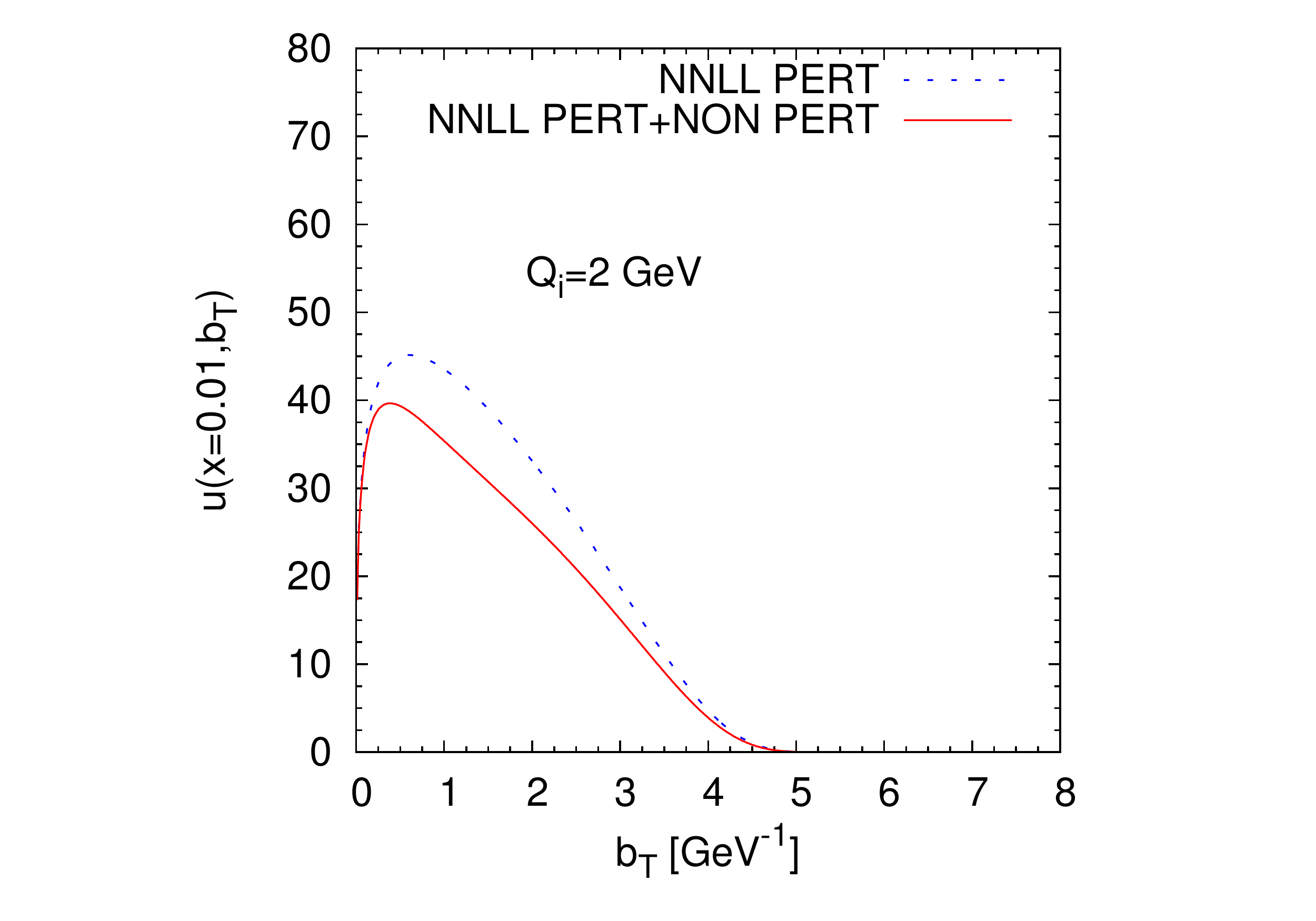}
 \caption{A comparison, in impact parameter space, between the perturbative part (blue-dashed line) of the TMDPDF for up quark at the initial scale $Q_i=2$ GeV, $x=0.01$ and the full TMDPDF (red-solid line), adopting the non-perturbative model as in Eq.~(\ref{eq:FqN3}) with the best-fit parameters $\lambda_1\sim 0.3$ and $\lambda_2\sim 0.13$.
\label{fig:Fpertn}}
 \end{center}
 \end{figure}

We anticipate here that the sensitivity of the data to this extra factor with $\l_3$ is not very  strong, although we observe an improvement in the $\chi^2$.
This is a consequence of the fact that the fully resummed $D$ function is actually valid on a region of impact parameter space which is broad enough for the analysis of the sets of available data (notice that we have in all cases a dilepton invariant mass $Q>4$ GeV). It might be that at lower values of $Q$ such corrections could be more significant. On the other hand one expects that also the factorization theorem should be revised when the values of $Q$ become of the order of the hadronization scale. It is then possible that the non-perturbative corrections to the evolution kernel happen there where the basic hypothesis of the factorization theorem ($Q\gg q_T\sim\Lambda_{QCD}\sim {\cal O}(1\;{\rm GeV})$) become weaker and so are more difficult to extract. A more detailed study in this direction is beyond the scope of this paper.

Finally in Fig.~\ref{fig:Fpertn} we show the effect of the model of Eq.~(\ref{eq:FqN3}) on the TMDPDF at low scale,  $Q=2$ GeV, where we expect that its impact is more substantial. We see that the non-perturbative correction affects mainly the intermediate $b_T$ region, while the high-$b_T$ region keeps naturally suppressed.

As a general remark one has to keep in mind that in practical calculation we finally Fourier transform the product of two TMDPDFs. The integration in impact parameter space is done numerically over a suitable $b_T$ range. We have checked that the region outside the endpoints of this integration does not affect the final result. In fact, the points for very small $b_T$ are relevant only for extremely high transverse momenta, which is not the case of our work. At very high $b_T$ the TMD are completely negligible (see Fig.~\ref{fig:Fpertn}).

To conclude this section we observe that while the parameter $\la_3$, being a correction to the $Q$-dependent piece of the TMD, is flavor independent, the other  parameters $\la_{1,2}$ can in principle be flavor dependent.
In the fit that we have performed we have not included this feature, namely for two reasons: $i)$ the DY data we use depend just on one combination of $\l_{1,2}$ (remember that we consider neutral current mediated processes and only protons as initial states); $ii)$ the quality of the fit is so good that we would not be sensitive (statistically) to the flavor dependence of these parameters.
Nevertheless the inclusion of data from processes with different initial states and/or mediated by charged currents could definitely help in this respect.
This is beyond the scope of the present work and will be considered in the future.

The results for the global fit are detailed in the next sections.

\section{Phenomenological Analysis}
\label{pheno}
We now move to a comparison of our theoretical estimates based on the above approach to available measurements on unpolarized cross sections. We perform a fit on different sets of data at various energies and covering a large $q_T$ range. This will allow us to fix the parameters entering the non-perturbative model, test the scale evolution of TMDs and the validity of the whole approach and highlight its main features. We will also discuss and quantify the uncertainties on the parameters so extracted, coming both from the statistical analysis in the fitting procedure as well as from some theoretical assumptions of our calculations. We will start with the selection of data discussing their role in our fit and then present our results.

\subsection{Data}
\label{data sel}

The selection of experimental data used in our fit is an important issue. Definitely we need data at moderate center of mass energies covering the small-$q_T$ region (up to 1-2 GeV) and intermediate dilepton invariant mass values (below 10 GeV). These come mainly from fixed-target experiments.  On the other hand to access larger $q_T$ values and even larger scales we have to include also high-energy collider experimental data, like those from Tevatron at the $Z$ pick. In both cases we will keep fulfilling the requirement $q_T\ll Q$, region of application of our approach.
Notice that to conform with the standard notation adopted in experimental analysis in the following we use $M=Q$ for the dilepton invariant mass.

These two classes of data are indeed complementary and essential to test the scale evolution of TMDs over a suitable range of scale values and to quantify the role of the non-perturbative part entering these distributions.

More precisely we consider the following data sets (see also Tabs.~\ref{datalow} and \ref{datahigh}, where we collect further details):
\begin{itemize}
\item moderate energy $p$-Cu and $pp$ data, $0<q_T<$ 1-2 GeV, 4 GeV $<M<25$ GeV:
\begin{itemize}
\item E288 data~\cite{Ito:1980ev} at $\sqrt s = 19.4$, 23.8 and 27.4 GeV;
\item E605 data~\cite{Moreno:1990sf} at $\sqrt s = 38.8$ GeV;
\item R209 data~\cite{Antreasyan:1981uv} at $\sqrt s = 62$ GeV;
\end{itemize}
\item high-energy $p\bar p$ data, $0<q_T<20$ GeV, $M = M_Z$:
\begin{itemize}
 \item CDF Run I~\cite{Affolder:1999jh} at $\sqrt s = 1.8$ TeV;
 \item D0 Run I~\cite{Abbott:1999yd,Abbott:1999wk} at $\sqrt s =1.8$ TeV;
 \item CDF Run II~\cite{Aaltonen:2012fi} at $\sqrt s = 1.96$ TeV;
 \item D0 Run II~\cite{Abazov:2007ac} at $\sqrt s =1.96$ TeV\,.
\end{itemize}
\end{itemize}

\begin{table}[h]
\begin{center}
\renewcommand{\tabcolsep}{0.4pc} 
\renewcommand{\arraystretch}{1.2} 
\begin{tabular}{|c|c|c|c|c|c|}
 \hline
 ~                        &  E288 200    &  E288 300        &  E288 400          &  E605               &   R209   \\
 \hline
 points                   &      35      &   35             &       49           &     -               &   6 \\
 \hline
 $\sqrt{s}$               &    19.4 GeV   &   23.8 GeV        &      27.4 GeV    &  38.8 GeV           &    62 GeV \\
\hline
 $E_{\rm beam}$               &   200 GeV     &   300 GeV        &  400 GeV          &   800 GeV           &    - \\
 \hline
Beam/Target               &  $p$ Cu         & $p$ Cu             &  $p$ Cu             & $p$ Cu                & $p p$       \\
 \hline
$M$ range                 &  4-9 GeV      &  4-9 GeV         &  5-9, 11-14 GeV   &  7-9, 10.5-18 GeV   &  5-8, 11-25 GeV\\
 \hline
 kin. var.           & $y$=0.4         &  $y$=0.21          &   $y$=0.03         &    $-0.1<x_F< 0.2$  &        \\
 \hline
 Observable               & $E d^3 \sigma/d^3\boldsymbol{q}$& $E d^3 \sigma/d^3\boldsymbol{q}$&$E d^3 \sigma/d^3\boldsymbol{q}$& $E d^3 \sigma/d^3\boldsymbol{q}$& $d\sigma/d q_T^2 $\\
\hline
 \end{tabular}
 \caption{Low-energy Drell-Yan experiments~\cite{Ito:1980ev,Moreno:1990sf,Antreasyan:1981uv}: numbers of points, center of mass and beam energy, beams and targets, invariant mass bins, fixed or integrated kinematical variables, and observables considered.  \label{datalow}}
\end{center}
\end{table}

\begin{table}[h]
\begin{center}
\renewcommand{\tabcolsep}{0.4pc} 
\renewcommand{\arraystretch}{1.2} 
\begin{tabular}{|c|c|c|c|c|}
 \hline
 ~                        & CDF Run I    &  D0 Run I        & CDF Run II        & D0 Run II      \\
 \hline
 points                   &      32      &   16             &       41          &        9       \\
 \hline
 $\sqrt{s}$               &      1.8 TeV &   1.8 TeV        &       1.96 TeV    &       1.96 TeV   \\
\hline
 $\sigma_{\rm exp}$       &$248\pm11$ pb & $221\pm 11.2$ pb &  $256\pm 15.2$ pb &    $255.8\pm16.7$ pb      \\
 \hline
$\sigma_{\rm teo}$ NNLO   & 224.6 pb     &     224.6 pb     &       246.1 pb    &        246.1  pb     \\
 \hline
Ratio exp/NNLO         &  1.104       &    0.98          &       1.04        &       1.04    \\
 \hline
$\sigma_{\rm teo}$ NLO      & 222.7 pb     &     222.7 pb     &       244.0 pb         &        244.0  pb     \\
 \hline
Ratio exp/NLO         &  1.114       &    0.992          &       1.049        &       1.048    \\
\hline
 Observable               & $(1/\sigma) d \sigma/d q_T$ & $(1/\sigma) d \sigma/d q_T$ & $(1/\sigma) d \sigma/d q_T$ & $(1/\sigma) d \sigma/d q_T$\\
\hline
\end{tabular}
\caption{High-energy data sets~\cite{Affolder:1999jh,Abbott:1999yd,Abbott:1999wk, Aaltonen:2012fi, Abazov:2007ac,  Abbott:1999tt,Abulencia:2005ix}: number of points, center of mass energy, experimental values of the total cross section, its theoretical estimates at NNLO and NLO (based on DYNNLO code by Catani \emph{et al.}~\cite{Catani:2007vq,Catani:2009sm}) with their corresponding ratios to the measured values, and the observable considered.\label{datahigh}}
\end{center}
\end{table}

While for the low-energy data we consider the invariant differential cross section in the virtual boson momentum, for the high-energy data sets we use the ratio of their $q_T$ dilepton distribution normalized to the experimental total cross section. In such a case, we compute this numerator following our approach and use the normalization factor as obtained with the DYNNLO code of Catani \emph{et al.}~\cite{Catani:2007vq,Catani:2009sm}. The use of this ratio avoids the problem of the discrepancy between D0 and CDF experiments (see Tab.~\ref{datahigh}) that could cause a source of systematics and/or tension between data sets.

As a first attempt of our global fit we include all data shown in Tables~\ref{datalow} and \ref{datahigh}. From this we realize that the E605 data set gives a very large, and anomalous, contribution to the total $\chi^2$. We then decide to study separately and in some detail the role played by these data. This is the outcome of this dedicated analysis:
\begin{itemize}
\item a separate fit of these data alone gives a large $\chi^2$ by itself ($\chi^2_{\rm points}\simeq 10$). The bad overall $\chi^2$ is not simply a matter of tension between these data and the others;
\item the different invariant mass bins seem not to be consistent among them and not to follow a proper scale dependence (even considering a $Q$-dependent non-perturbative part in the TMD);
\item the parameters resulting from the global fit are almost unaffected by the inclusion of these data;
\item in the literature the inclusion of this data set seems also problematic. In Ref.~\cite{Landry:2002ix} only the two bins at the lowest dilepton invariant masses are included, without any comment on the rest of data.
\end{itemize}

{}For these reasons we prefer to exclude the E605 data set in the following global analysis. We will further comment on the role of these data when comparing our study with other fits, Section~\ref{sec:lit}.

\subsection{Fitting procedure and results}
\label{fit}

We perform different fits to test various aspects of our approach. We start by considering only the high-energy data set from Tevatron~\cite{Affolder:1999jh,Abbott:1999yd,Abbott:1999wk, Aaltonen:2012fi, Abazov:2007ac}, adopting for the non-perturbative factor, $\tilde F^{\rm NP}$ in Eq.~(\ref{eq:FqN3}), the exponential part alone without the correcting power-like behaviour. This results in a very good fit ($\chi^2_{\rm d.o.f.}= 0.44$) showing how the model is well sound and that a one-parameter fit is enough to describe the $Z$-boson spectrum. We also tried a Gaussian functional form - that implies a Gaussian form also in momentum space (as commonly adopted in many parametrizations of TMDs) - but the resulting $\chi^2$ is worse.

The use of the exponential form, instead of the more usual Gaussian-like form, has been suggested in Refs.~\cite{Schweitzer:2012hh} and~\cite{Collins:2013zsa} based on the fact that Euclidean correlation functions in quantum field theory are usually exponential and not Gaussian at large distances. To our knowledge it is the first time that this hypothesis has been tested explicitly on a phenomenological analysis and within the TMD formalism.

This preliminary study allows us to draw some interesting conclusions:
\begin{itemize}
\item the large-$q_T$ spectrum ($q_T$ above the pick, say $q_T\gtrsim 5$ GeV) at Tevatron is very well reproduced by the perturbative, and therefore calculable, part of the TMDs. This is a non trivial result showing the consistency of the approach as well as the resummation procedure.
\item The non-perturbative piece is necessary to describe the data points below and around the pick, with a strong preference in favor of an exponential damping instead of a Gaussian shape in $b_T$ space. As the data are at fixed dilepton invariant mass ($M_Z$) one cannot check the $Q$-dependence of the non-perturbative model considering this set of data alone.
\item Being the low-$q_T$ region populated by very few data points, to which the non-perturbative part should be more sensitive, our fit at NNLL appears somehow over-determined with a $\chi^2_{\rm d.o.f.}\ll 1$. Notice that at NLL accuracy this is not the case, showing once again the importance of a NNLL analysis.
\end{itemize}

As a second step we include also lower-energies DY data and perform a global fit. This fit, as already pointed out, covers the very important small-$q_T$ region and therefore is expected to strongly constrain the non-perturbative piece. This is what we will discuss in the following.

Differently from the Tevatron data, the low-energy data are affected  by large uncertainties in the normalization of the cross section. For this reason we allow for two extra parameters in the fit, namely the normalization factors for E288 and R209 data: $N_{\rm E288}$ and $N_{\rm R209}$. Notice that these extra two parameters are strongly related to the present accuracy of the available experimental data, while the two (three) parameters entering our TMDPDF, $\lambda_1, \,\lambda_2\, (\l_3)$ are the main goal of our analysis.

The global fit of Tevatron and low-energy data adopting a simple exponential factor in the non-perturbative  model is very bad and the description of the Tevatron data results somehow spoiled. This result does not change if we allow a $Q$ dependence of the non-perturbative model, alike  in Eq.~(\ref{eq:FqN3m}).
This is a signal that the low-$q_T$ region, so sensitive to the large-$b_T$ behaviour of the TMD, requires more care.
For this reason we exploit various simple extensions of the functional form of $\tilde F^{\rm NP}$ and end up with the power-like piece (quadratic in $b_T$) entering  Eqs.~(\ref{eq:FqN3}), (\ref{eq:FqN3m}): this is what better describes the data.

We perform a fit both at NLL and NNLL. When adopting the NLL approximation we use the next-to-leading-order (NLO) collinear parton distributions, while at NNLL we use the NNLO PDFs. In both cases we adopt $Q_i=Q_0+q_T$ and for the collinear PDFs we use the MSTW08 set~\cite{Martin:2009iq}.
We also check the role played by the collinear PDFs adopting another set, namely the CTEQ10~\cite{Lai:2010vv} and we find a complete consistency among the results
(the uncertainty due to different choices of PDF sets can be deduced by comparing the tables in the text and in Appendix~\ref{app:cteq}).

One of the main goals of this work consists in the fits performed at NNLL accuracy with full resummation.
The NLL fits are mainly used as a check of convergence of the theory and other phenomenological aspects. We have tested both a $Q$-independent and $Q$-dependent parametrization of the non-perturbative inputs as given respectively in Eq.~(\ref{eq:FqN3}) and Eq.~(\ref{eq:FqN3m}). The results corresponding to the model in Eq.~(\ref{eq:FqN3}) are  summarized in Tables~\ref{tab:tevlow_mstw08_q0pt_points}-\ref{tab:tevlow_mstw08_q0pt_param} (and Tables~\ref{tab:CTEQ1}-\ref{tab:CTEQ2} for the CTEQ10 set of PDFs), while the ones corresponding to the model in Eq.~(\ref{eq:FqN3m}) are summarized in Tables~\ref{tab:tevlow_mstw08_q0pt_pointsQ} -\ref{tab:tevlow_mstw08_q0pt_paramQ} (and Table~\ref{tab:CTEQ3} for the CTEQ10 set of PDFs). We will describe in some detail both cases.

In the following the statistical error is estimated requiring a 68\% confidence level, corresponding to a $\Delta \chi^2=4.72$ for four parameters.
Concerning the theoretical errors (which include the error due to uncalculated perturbative terms still using the full resummation), we study the dependence on the initial scale $Q_i=Q_0+q_T$ (where $Q_0=2$ GeV) in two ways: $i)$ we check the impact of a change in $Q_0$ allowing $m_{\rm charm}\sim  1.3 \;{\rm GeV}\leq Q_0\leq 2.7$ GeV, where the lowest value of $Q_0$ is about the charm threshold and the highest value is limited by the energy of the lowest energy bin of data; $ii)$ keeping  $Q_0=2$ GeV, we vary $Q_0+q_T/2\leq Q_i\leq {\rm min}\;(Q_0+2 q_T, Q)$. The latter error is the one due to the residual scale dependence. In the next subsection we comment on the  impact of this scale variation on the fit of the parameters. The theoretical error due to the scale variation does not substantially depend on the used PDF set and so it is not reported in Tab.~\ref{tab:CTEQ1}-\ref{tab:CTEQ3}. In Section~\ref{sec:scakes} we check the stability of the result due to the scale variation.

\subsubsection*{$Q$-independent non-perturbative input ($D^{\rm NP}=0$)}

Adopting the model in Eq.~(\ref{eq:FqN3}) the data at our disposal can be described with a $\chi^2/{\rm d.o.f.}\sim 1$, both at NLL and at NNLL as shown in Tables~\ref{tab:tevlow_mstw08_q0pt_points} and \ref{tab:tevlow_mstw08_q0pt_param}  (and Tables~\ref{tab:CTEQ1}-\ref{tab:CTEQ2} for the CTEQ10 set of PDFs).

The NNLL-NNLO calculation gives a better overall $\chi^2$ and definitely improves the description of Tevatron data. Concerning the low-energy data, while the E288 fit is practically unaffected, for R209 we have a strong correlation between the normalization factor and the relative $\chi^2$ that eventually gives a good description in both approximations.
By looking at Tab.~\ref{tab:tevlow_mstw08_q0pt_points} we see that for each high-energy data set, that means also large-$q_T$ and large-$Q$ values, the $\chi^2$ for data points improves significantly going from NLL to NNLL accuracy. This is somehow expected since in this region the logarithmic terms are more important. On the other hand, using the MSTW08 set for the PDFs, we slightly loose in two low-energy data sets, still achieving a better overall $\chi^2$. Notice that adopting the PDFs from CTEQ10 we have an improvement of the $\chi^2$ for all data samples (see Tab.~\ref{tab:CTEQ1}).

\begin{table}[ht]
\begin{tabular}{|c|c|c|c|c|c|}
 \hline
      ~        &   ~        &  NNLL, NNLO       &    NLL, NLO \\
 \hline
     ~        &    points  &   $\chi^2/\textrm{points}$  &    $\chi^2/\textrm{points}$   \\
 \hline
              &    223     &      1.10          &     1.48           \\
 \hline
 \hline
 E288 200     &     35    &      1.53            &    2.60      \\
  \hline
 E288 300     &     35     &       1.50          &    1.12       \\
 \hline
 E288 400     &     49     &       2.07          &    1.79     \\
 \hline
 \hline
 R209         &      6    &       0.16           &     0.25            \\
 \hline
 \hline
  CDF Run I   &     32    &       0.74           &     1.31                \\
 \hline
  D0 Run I    &     16     &       0.43          &     1.44      \\
 \hline
  CDF Run II  &     41     &       0.30          &     0.62    \\
 \hline
  D0 Run II   &      9     &       0.61          &     2.40   \\
 \hline
\end{tabular}
\caption{Total and partial $\chi^2/\textrm{points}$ of our global fit on low-energy~\cite{Ito:1980ev,Antreasyan:1981uv} and Tevatron data~\cite{Affolder:1999jh,Abbott:1999yd,Abbott:1999wk, Aaltonen:2012fi, Abazov:2007ac} with $D^{\rm NP}=0$ (Eq.~(\ref{eq:FqN3})), \mbox{$Q_i=Q_0+q_T$}, at NNLL and NNL accuracies and with the collinear parton distributions from MSTW08~\cite{Martin:2009iq} at NNLO and NLO. \label{tab:tevlow_mstw08_q0pt_points}}
\end{table}

\begin{table}[ht]
\renewcommand{\tabcolsep}{0.4pc} 
\renewcommand{\arraystretch}{1.6} 
\begin{center}
\begin{tabular}{|l|l|l|}
 \hline
          NLL                       &  223  points           &   $\chi^2$/d.o.f. = 1.51  \\
  \hline
 ~                                  &  $\lambda_1$ = $0.26^{+0.05_{\rm th}}_{-0.02_{\rm th}}\pm0.05_{\rm stat}\textrm{ GeV}$ &   $\lambda_2=0.13\pm0.01_{\rm th}\pm0.03_{\rm stat}\textrm{ GeV}^2$  \\
\hline
~                                   &  $N_{\rm E288}=0.9^{+0.2_{\rm th}}_{-0.1_{\rm th}}\pm0.04_{\rm stat}$&$N_{\rm R209}=1.3\pm0.01_{\rm th}\pm0.2_{\rm stat}$\\
\hline
\hline
          NNLL                       &  223  points           &   $\chi^2$/d.o.f. = 1.12  \\
  \hline
 ~                                  &  $\lambda_1$ = $0.33\pm0.02_{\rm th}\pm0.05_{\rm stat}\textrm{ GeV}$ &   $\lambda_2=0.13\pm0.01_{\rm th}\pm0.03_{\rm stat}\textrm{ GeV}^2$  \\
\hline
~                                   &  $N_{\rm E288}=0.85\pm0.01_{\rm th}\pm0.04_{\rm stat}$&$N_{\rm R209}=1.5\pm0.01_{\rm th}\pm0.2_{\rm stat}$\\
\hline
\end{tabular}
\caption{Results of our global fit on low-energy~\cite{Ito:1980ev,Antreasyan:1981uv} and Tevatron data~\cite{Affolder:1999jh,Abbott:1999yd,Abbott:1999wk, Aaltonen:2012fi, Abazov:2007ac}, with $D^{\rm NP}=0$ (Eq.~(\ref{eq:FqN3})), \mbox{$Q_i=Q_0+q_T$}, at NNLL and NNL accuracies and with the collinear parton distributions from MSTW08~\cite{Martin:2009iq} at NNLO and NLO. \label{tab:tevlow_mstw08_q0pt_param}}
\end{center}
\end{table}

\begin{figure}[h!t]
 \begin{center}
 \includegraphics[width=0.45\textwidth, angle=0,natwidth=610,natheight=642]{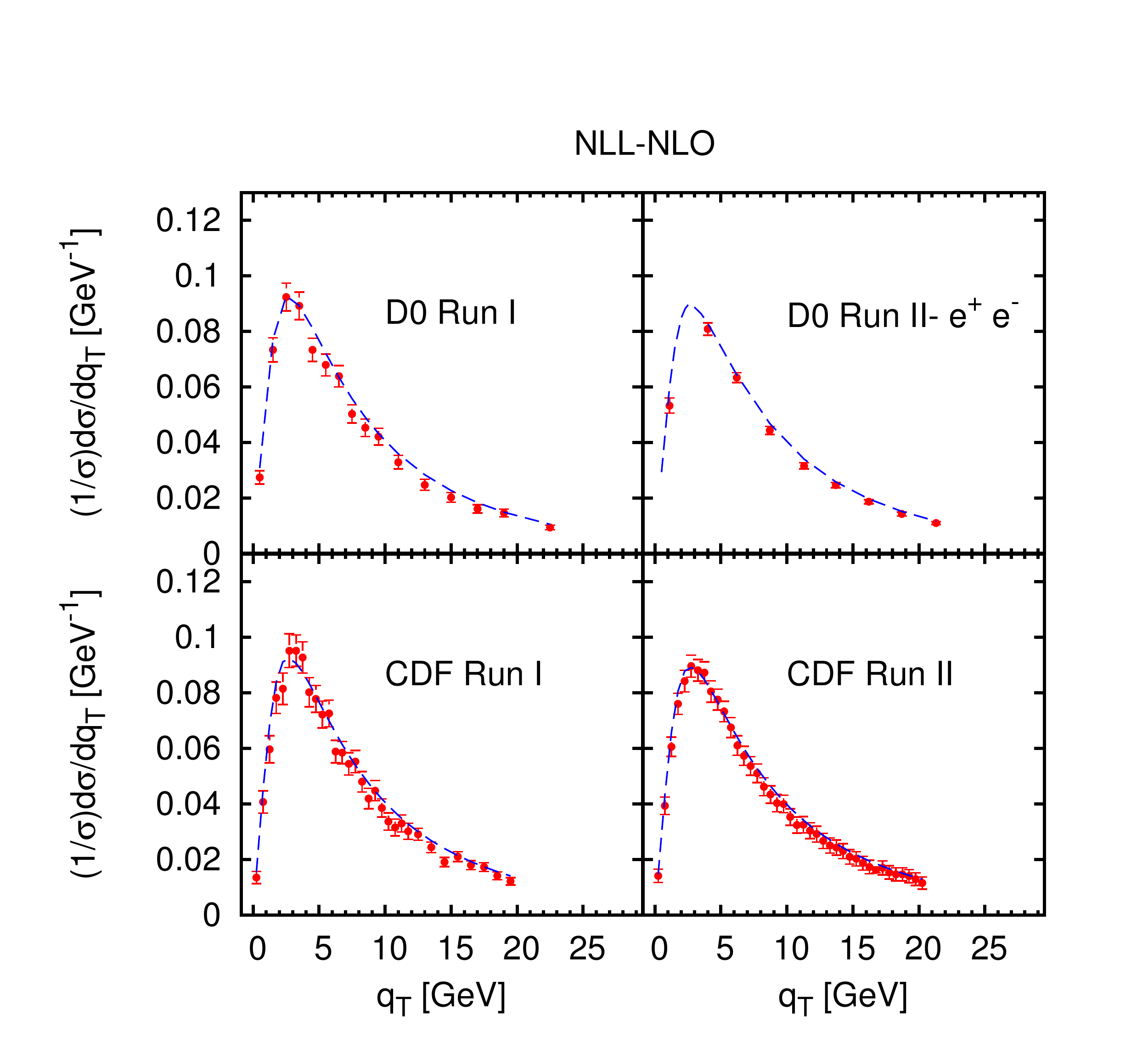}
\includegraphics[width=0.45\textwidth, angle=0,natwidth=610,natheight=642]{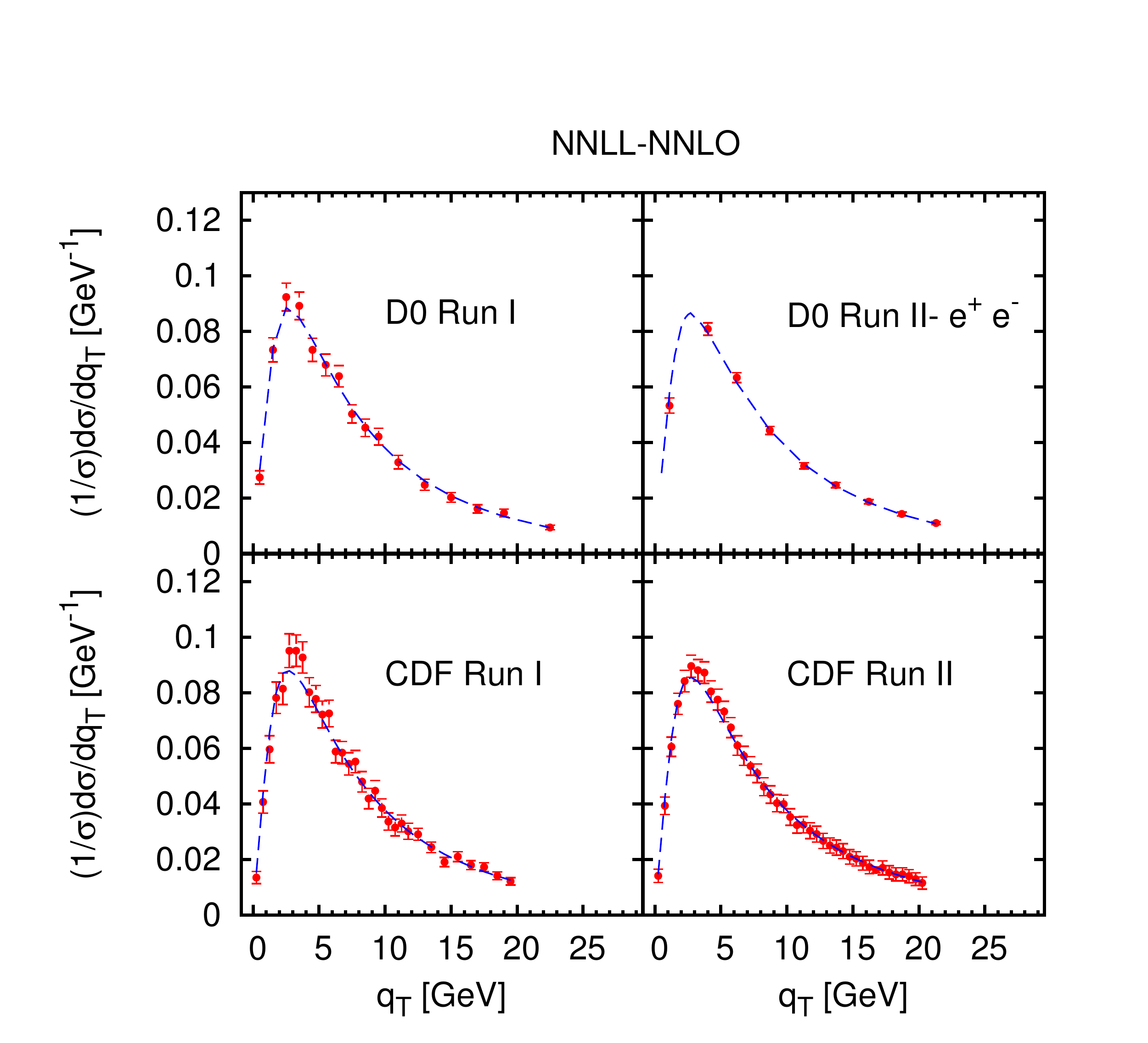}\
 \caption{Comparison of our theoretical estimates for $(1/\sigma)\, d\sigma/dq_T$ with Tevatron data~\cite{Affolder:1999jh,Abbott:1999yd,Abbott:1999wk, Aaltonen:2012fi, Abazov:2007ac}. The results are obtained from the global fit with $D^{\rm NP}=0$ (Eq.~(\ref{eq:FqN3})), $Q_i=Q_0+q_T$, at NLL accuracy  with collinear PDFs at NLO (left panel), and NNLL accuracy with NNLO PDFs (right panel). For the collinear PDFs we use the MSTW08 set~\cite{Martin:2009iq}.
\label{fig:high}}
 \end{center}
 \end{figure}

\begin{figure}[h!]
 \begin{center}
 \includegraphics[width=.8\textwidth, angle=0,natwidth=610,natheight=642]{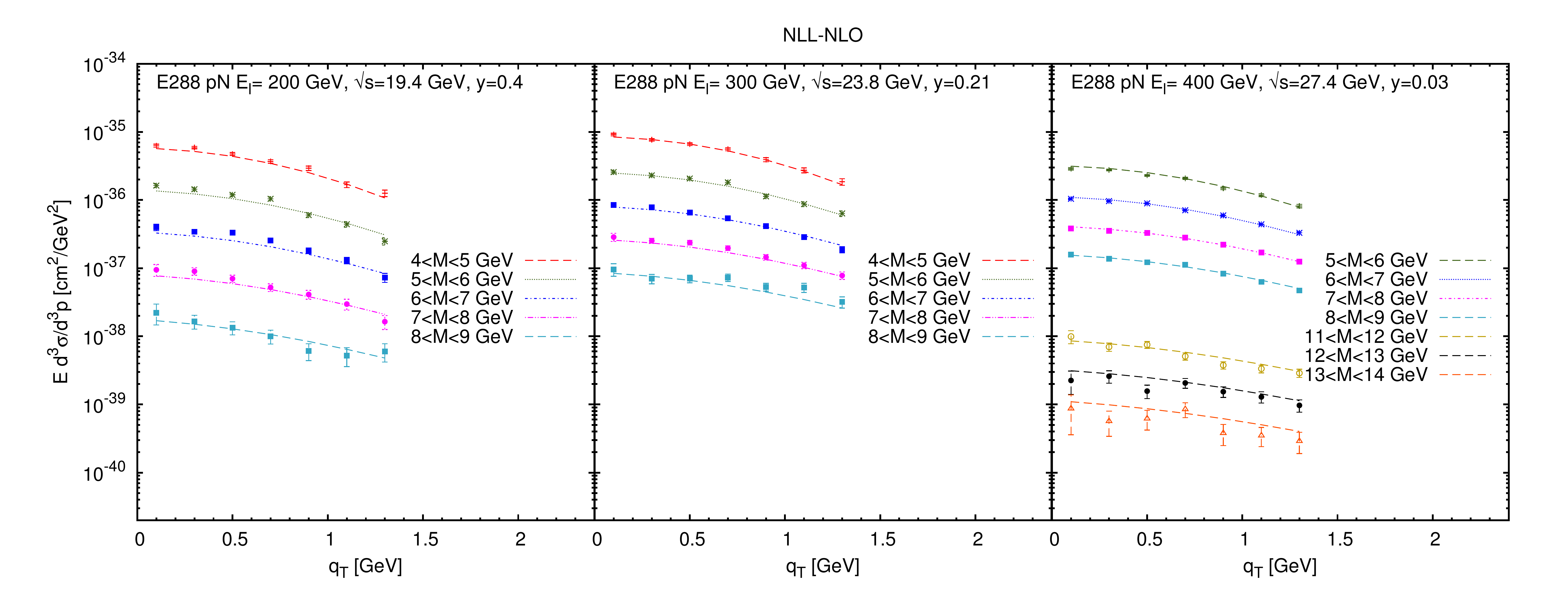}\\
\includegraphics[width=.8\textwidth, angle=0,natwidth=610,natheight=642]{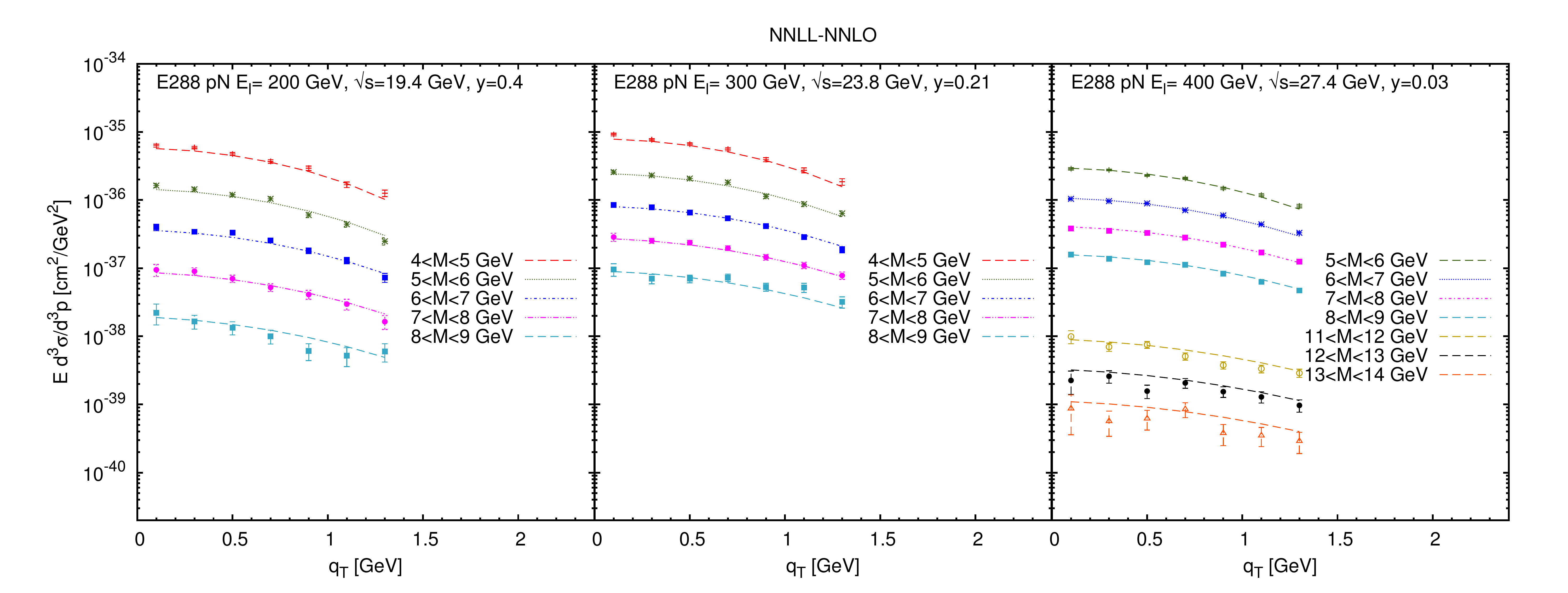}
 \caption{Comparison of our theoretical estimates for $d^3\sigma/d^3\bm{q}$ with E288 at three different energies~\cite{Ito:1980ev}. The results are obtained from the global fit with $D^{\rm NP}=0$ (Eq.~(\ref{eq:FqN3})), $Q_i=Q_0+q_T$, at NLL accuracy with collinear PDFs at NLO (upper panels) and at NNLL accuracy with NNLO PDFs (lower panels). For the collinear PDFs we use the MSTW08 set~\cite{Martin:2009iq}.
\label{fig:lowE288}}
 \end{center}
 \end{figure}

 \begin{figure}[h]
 \begin{center}
  \includegraphics[width=0.4\textwidth, angle=0,natwidth=610,natheight=642]{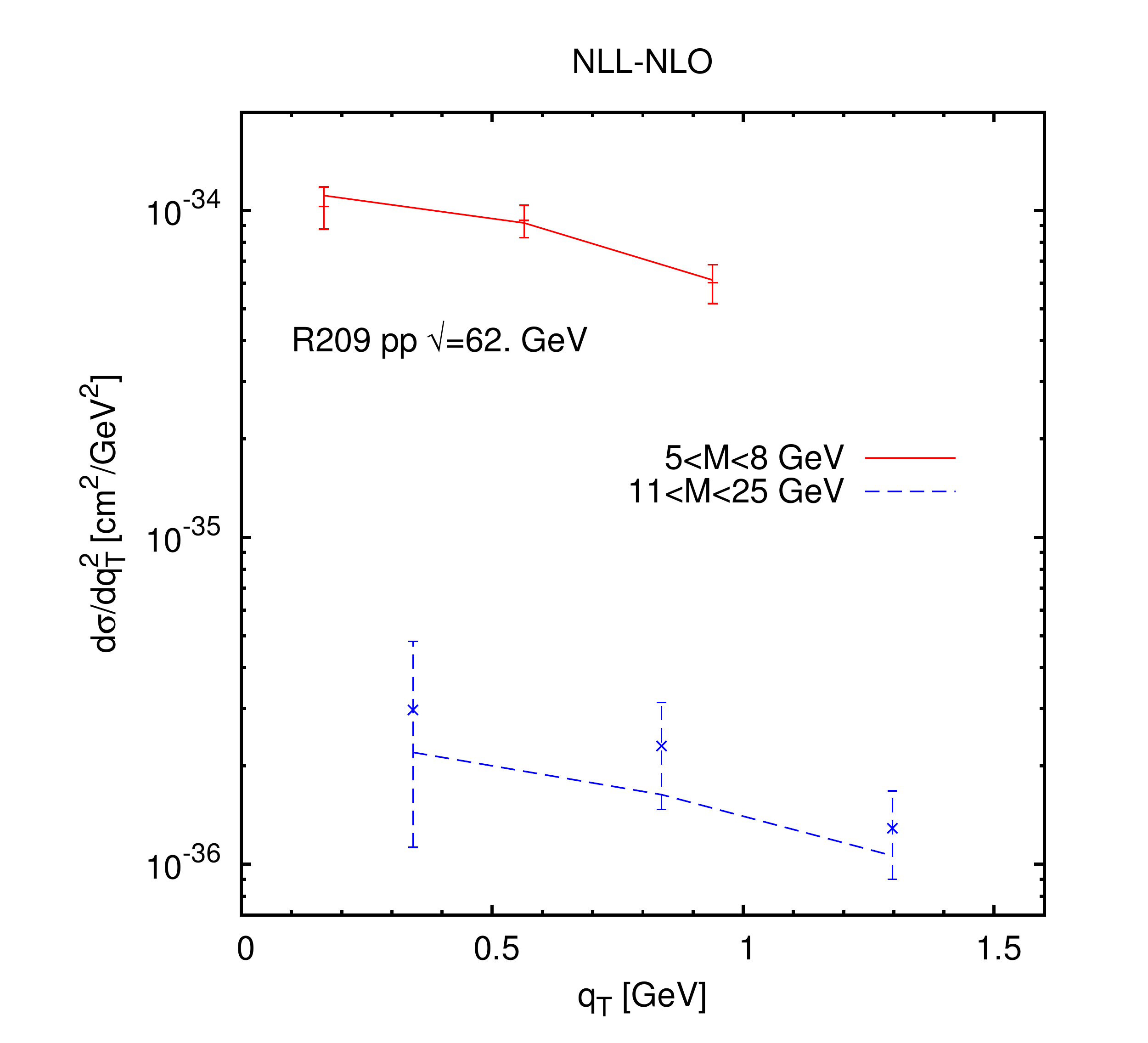}
  \includegraphics[width=0.4\textwidth, angle=0,natwidth=610,natheight=642]{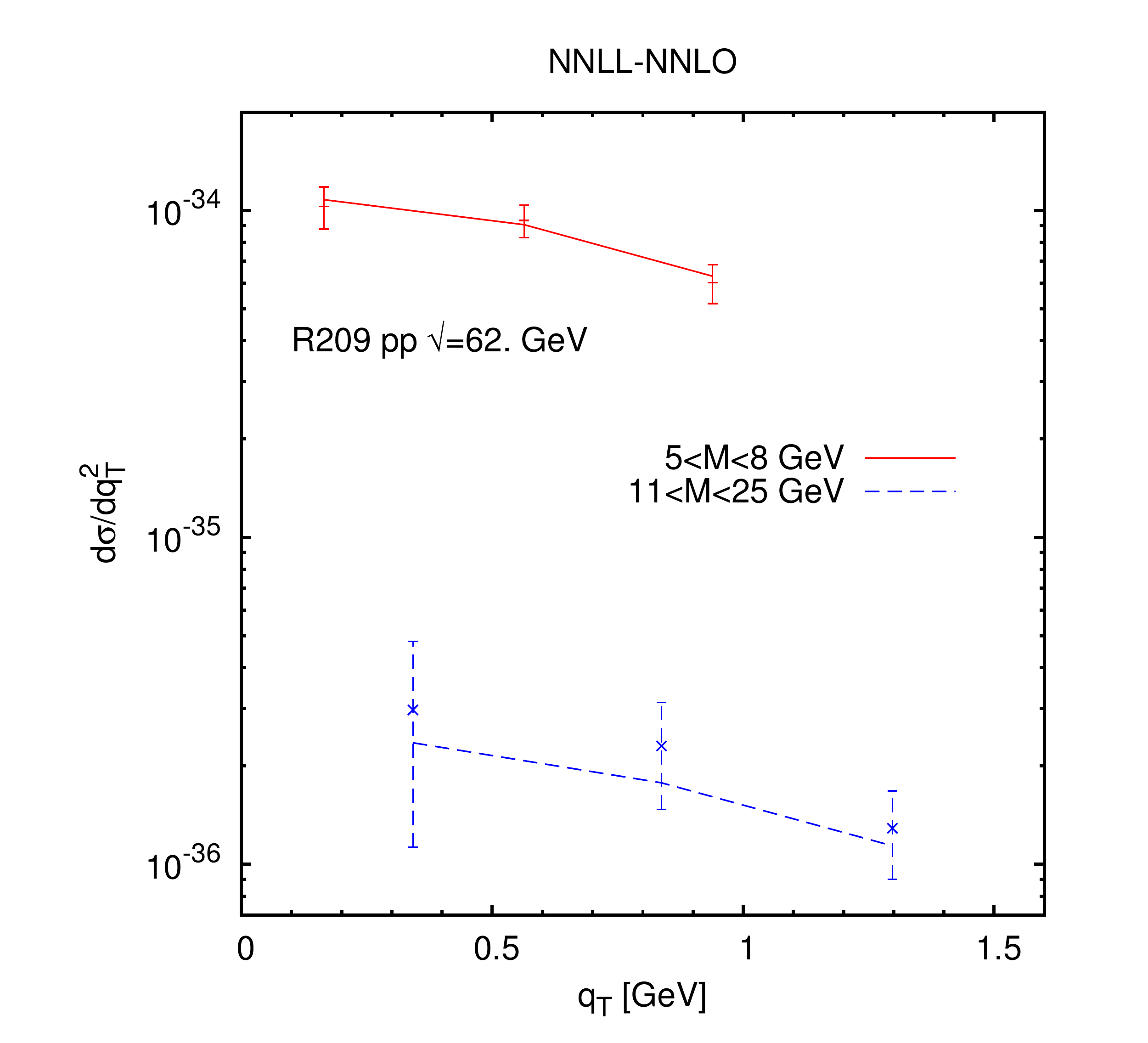}
 \caption{Comparison of our theoretical estimates for $d\sigma/dq_T^2$ with R209 data~\cite{Antreasyan:1981uv}. The results are obtained from the global fit with $D^{\rm NP}=0$ (Eq.~(\ref{eq:FqN3})), $Q_i=Q_0+q_T$, at NLL accuracy with collinear PDFs at NLO (left panel) and at NNLL accuracy with NNLO PDFs (right  panel). For the collinear PDFs we use the MSTW08 set~\cite{Martin:2009iq}.
\label{fig:lowR209}}
 \end{center}
 \end{figure}

\begin{figure}[h!t]
\vspace*{-3cm}
 \begin{center}
 \includegraphics[width=0.45\textwidth, angle=0]{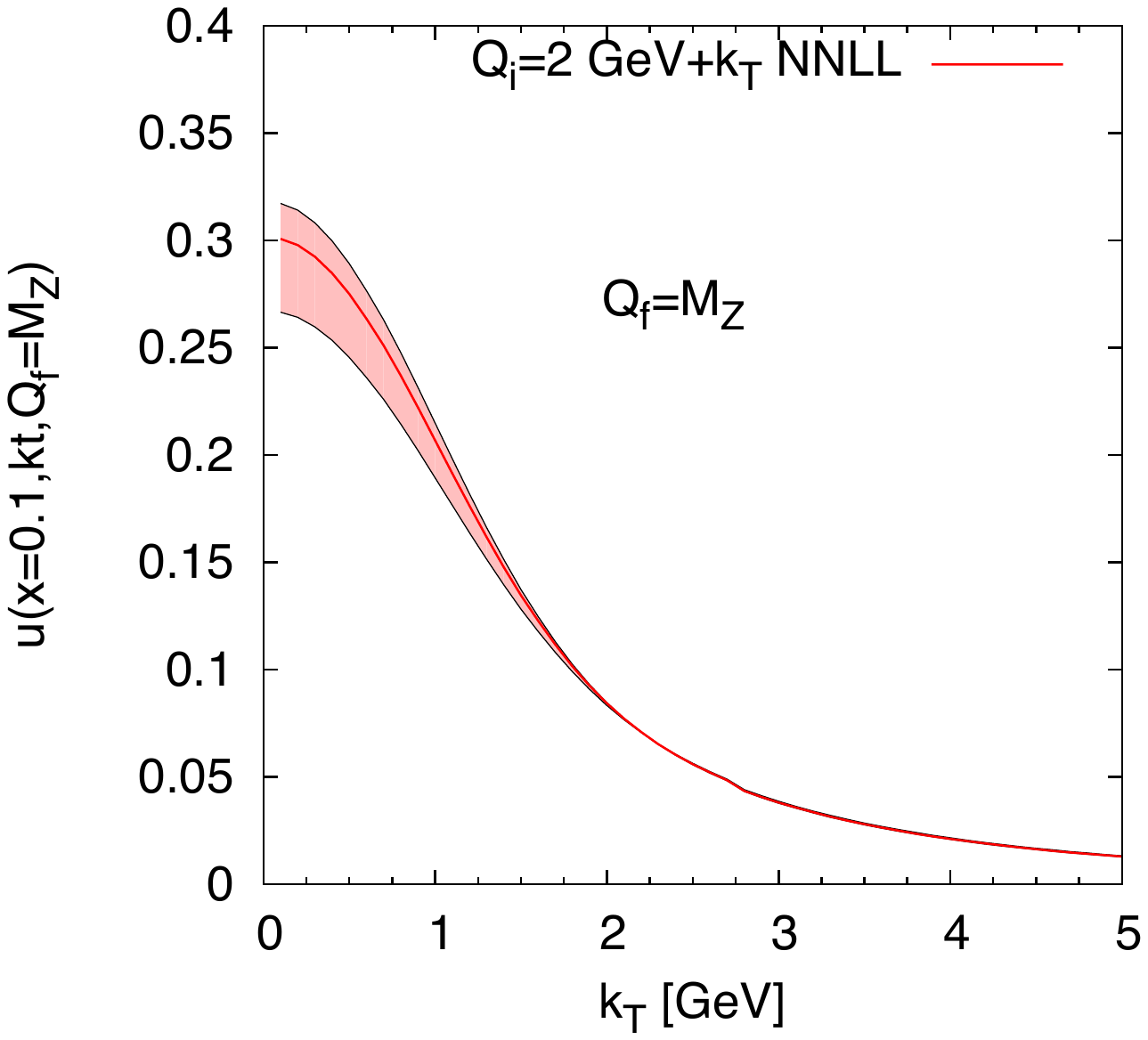}\hspace{-2cm}
\includegraphics[width=0.45\textwidth, angle=0,natwidth=610,natheight=642]{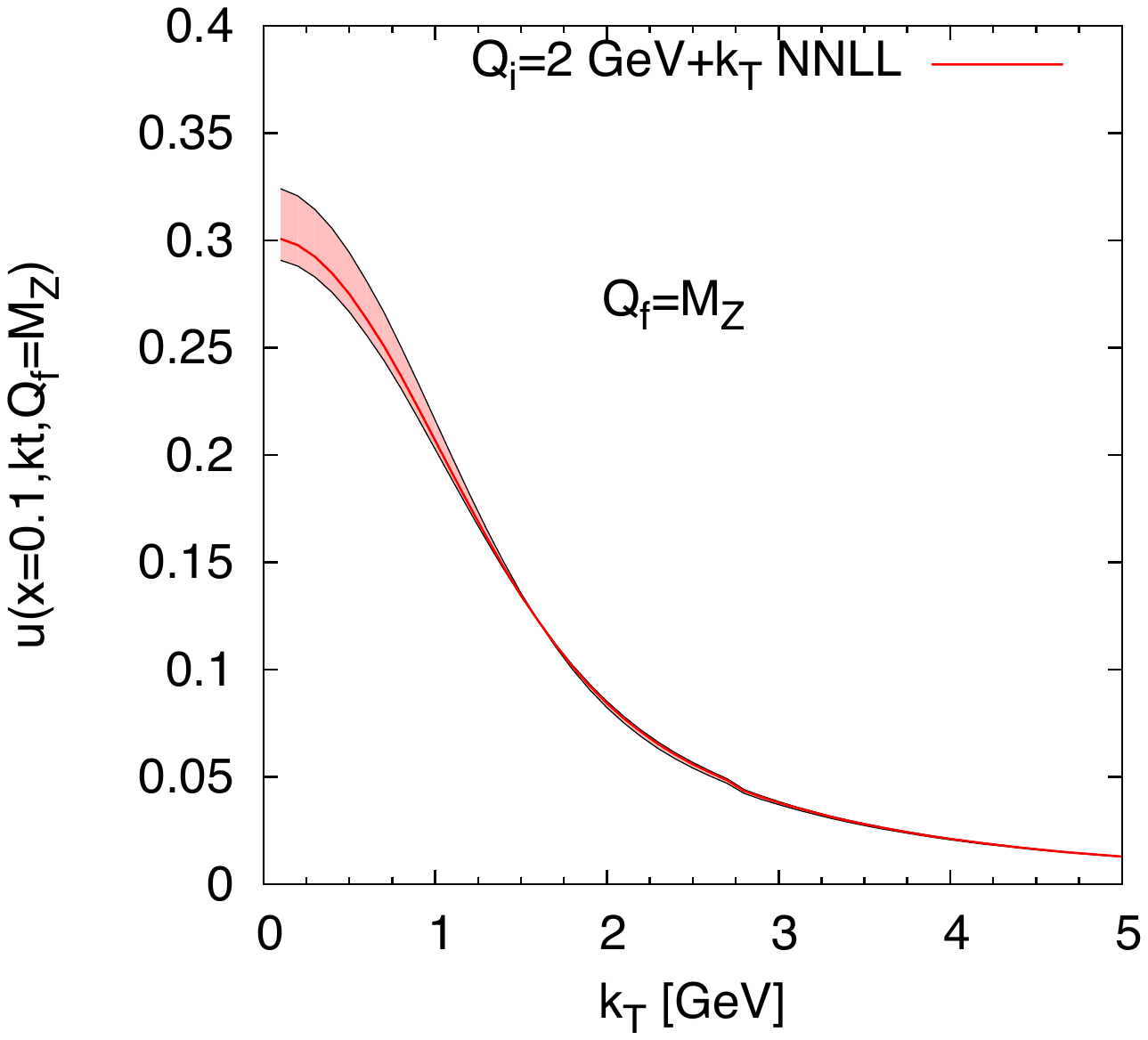}\\
\vspace{-6.5cm}
 \includegraphics[width=0.45\textwidth, angle=0,natwidth=610,natheight=642]{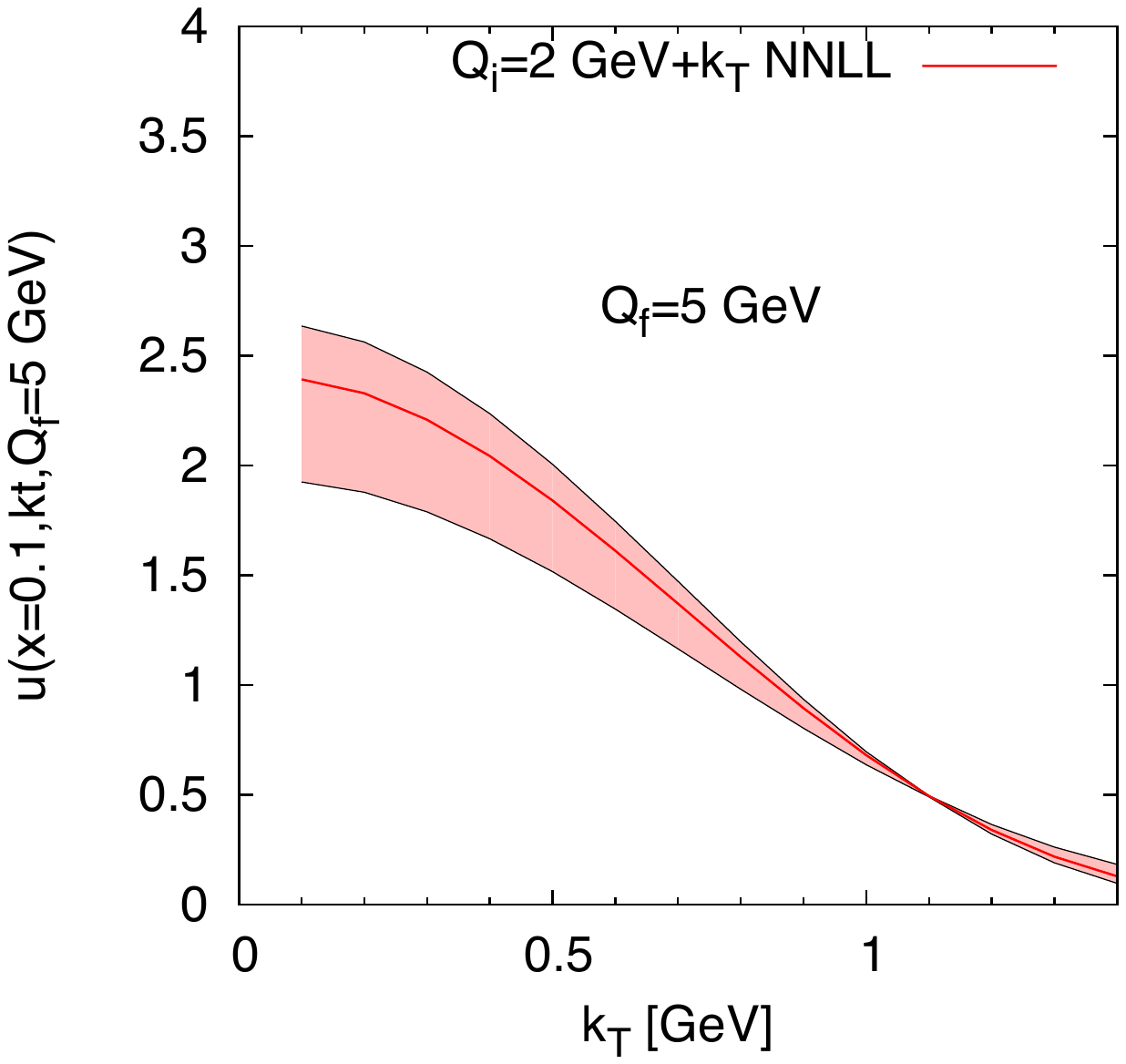}\hspace{-2cm}
\includegraphics[width=0.45\textwidth, angle=0,natwidth=610,natheight=642]{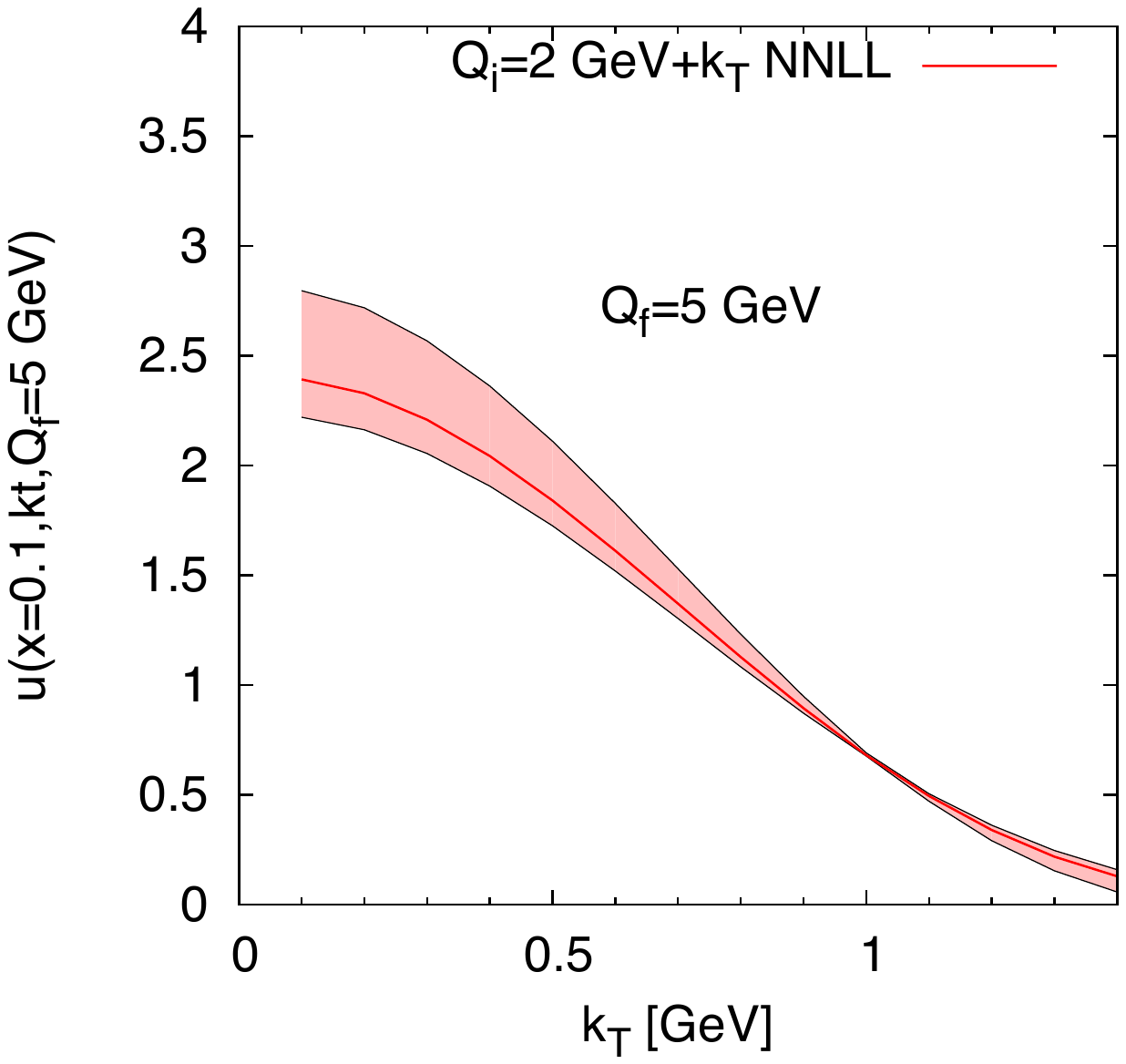}
\vspace{-3 cm}
 \caption{TMDPDF for up quark in momentum space at $x=0.1$ and $Q=Q_f=M_Z$ (upper plots) and $Q=Q_f=5$ GeV (lower plots) at NNLL. The bands are generated taking into account the statistical error on $\l_1$ (left panels) and  $\l_2$ (right panels).
\label{fig:lQ5}}
 \end{center}
 \end{figure}

The normalization factor for E288 data is always within their experimental uncertainty, while for R209 it is a bit larger. One observes however that the central values for R209 data are not so accurate - we notice here there are no official tables from the Collaboration - so that one can expect a larger normalization error for these data.
The parameter $\lambda_2$ is unaffected by the higher order contributions being the same in both approximation, while $\lambda_1$ presents some differences even if within the relative errors.
The technical reason for this shift is the appearance at NNLL of the one-loop contribution of the coefficients $\hat C$ as outlined in Tab.~\ref{tab:resummation} and visible also in Fig.~\ref{fig:Fpert}.
We expect that higher order contributions on this coefficients would stabilize the result.

The change in the scale $Q_0$ leaves the fit practically unaffected and the values of the parameters $\l_1,\l_2$ are only slightly changed.
The variation of the renormalization scale instead has some impact on the these values. In particular from Tab.~\ref{tab:tevlow_mstw08_q0pt_param} one can see that at NLL the theoretical error is of the same order of the statistical one and that there is a clear reduction of the scale dependence at NNLL.
At this order the main uncertainty on the fitted parameters comes from the statistical error.
In Fig.~\ref{fig:high} we show the comparison of our theoretical estimates with Tevatron at NLL and
NNLL accuracies, while in Figs.~\ref{fig:lowE288} and \ref{fig:lowR209} we show the corresponding comparison for low-energy data. In Fig.~\ref{fig:lQ5} we present the impact of the statistical error on the TMDPDF at NNLL at $Q=M_Z$ and $Q=5$ GeV.


\begin{table}[th]
\begin{tabular}{|c|c|c|c|c|c|}
 \hline
      ~        &   ~        &  NNLL, NNLO       &    NLL, NLO \\
 \hline
     ~        &    points  &   $\chi^2/\textrm{points}$  &    $\chi^2/\textrm{points}$   \\
 \hline
              &    223     &      0.79          &     1.40          \\
 \hline
 \hline
 E288 200     &     35    &      1.24            &    2.27      \\
  \hline
 E288 300     &     35     &       0.90          &    1.20       \\
 \hline
 E288 400     &     49     &       1.33          &    1.69     \\
 \hline
 \hline
 R209         &      6    &       0.24           &     0.30            \\
 \hline
 \hline
  CDF Run I   &     32    &       0.58           &     1.26                \\
 \hline
  D0 Run I    &     16     &       0.36          &     1.43      \\
 \hline
  CDF Run II  &     41     &       0.15          &     0.48    \\
 \hline
  D0 Run II   &      9     &       0.36          &     2.26   \\
 \hline
\end{tabular}
\caption{Total and partial $\chi^2/\textrm{points}$ of our global fit on low-energy~\cite{Ito:1980ev,Antreasyan:1981uv} and Tevatron data~\cite{Affolder:1999jh,Abbott:1999yd,Abbott:1999wk, Aaltonen:2012fi, Abazov:2007ac} with $D^{\rm NP}\ne0$ (Eq.~(\ref{eq:FqN3m})), \mbox{$Q_i=Q_0+q_T$}, at NNLL and NNL accuracies and with the collinear parton distributions from MSTW08~\cite{Martin:2009iq} at NNLO and NLO. \label{tab:tevlow_mstw08_q0pt_pointsQ}}
\end{table}

\begin{table}[h]
\renewcommand{\tabcolsep}{0.4pc} 
\renewcommand{\arraystretch}{1.6} 
\begin{center}
\begin{tabular}{|l|l|l|}
 \hline
          NLL          &  223  points           &   $\chi^2$/d.o.f. = 1.44  \\
  \hline
 ~                     &  $\lambda_1$ = $0.24^{+0.06_{\rm th}}_{-0.02_{\rm th}} \pm 0.05_{\rm stat}\textrm{ GeV}$
                       &  $\lambda_2=0.17 \pm 0.02_{\rm th} \pm 0.05_{\rm stat}\textrm{ GeV}^2$  \\
\hline
~                      &  $\lambda_3=0.03 \pm 0.02_{\rm th} \pm 0.01_{\rm stat} \textrm{ GeV}^2$                                                             &~\\
\hline
~                      &  $N_{\rm E288}=0.85^{+0.2_{\rm th}}_{-0.1_{\rm th}} \pm 0.04_{\rm stat}$
                       &  $N_{\rm R209}=1.2  \pm 0.2_{\rm th} \pm 0.2_{\rm stat}$ \\
\hline
\hline
          NNLL         &  223  points           &   $\chi^2$/d.o.f. = 0.81  \\
  \hline
 ~                     &  $\lambda_1$ = $0.30 \pm 0.02_{\rm th} \pm 0.05_{\rm stat}\textrm{ GeV}$
                       &  $\lambda_2=0.22 \pm 0.01_{\rm th} \pm 0.05_{\rm stat}\textrm{ GeV}^2$  \\
\hline
~                      &  $\lambda_3=0.05 \pm 0.01_{\rm th} \pm 0.02_{\rm stat}\textrm{ GeV}^2$                                                             &~\\
\hline
~                      &  $N_{\rm E288}=0.78^{+0.08_{\rm th}}_{-0.04_{\rm th}} \pm 0.05_{\rm stat}$
                       &  $N_{\rm R209}=1.3\pm0.1_{\rm th} \pm 0.2_{\rm stat}$\\
\hline
\end{tabular}
\caption{Results of our global fit on low-energy~\cite{Ito:1980ev,Antreasyan:1981uv} and Tevatron data~\cite{Affolder:1999jh,Abbott:1999yd,Abbott:1999wk, Aaltonen:2012fi, Abazov:2007ac}, with $D^{\rm NP}\ne0$ (Eq.~(\ref{eq:FqN3m})), \mbox{$Q_i=Q_0+q_T$}, at NNLL and NNL accuracies and with the collinear parton distributions from MSTW08~\cite{Martin:2009iq} at NNLO and NLO.
 \label{tab:tevlow_mstw08_q0pt_paramQ}}
\end{center}
\end{table}

\subsubsection*{$Q$-dependent non-perturbative input ($D^{\rm NP}\ne 0$)}

The study of the $Q$ dependence of the fit within the model of Eq.~(\ref{eq:FqN3m}) is summarized in Tables~\ref{tab:tevlow_mstw08_q0pt_pointsQ} and \ref{tab:tevlow_mstw08_q0pt_paramQ} (and Table~\ref{tab:CTEQ3} for the CTEQ10 set of PDFs). It is important to notice that the $Q$-dependent term cannot substitute any other piece of the fit. In other words putting $\l_1=0$ and/or $\l_2=0$ and leaving only the term with $\l_3\neq 0$ in the model of Eq.~(\ref{eq:FqN3m}) does not produce a reasonable fit. The estimates of the error in the the fit and the parameters are done using the same criterium (68$\%$ confidence level) as in the case $\l_3=0$.

With a three-parameter fit we have an improvement of the total $\chi^2$, specially at NNLL. In fact, at NLL the uncertainties somehow mask the benefits of the introduction of a new correction and we do not obtain a significative change of the $\chi^2$. The core of the improvement in the $\chi^2$ is that all low-energy experiments are much better described at NNLL, while in the case of $\l_3=0$ the change from NLL to NNLL is more controversial for this data subset.
In other words, while the introduction of a new parameter in the fit is not conclusive from a NLL analysis, it seems instead more appropriate when all perturbative pieces are developed at NNLL. The final $\chi^2/{\rm d.o.f.}\sim 0.8$ at NNLL confirms that this kind of precision is necessary to extract important information on the non-perturbative structure of TMDs, that a NLL analysis may hide.

Comparing the values of $\l_{1,2}$ with and without the inclusion of the $Q$-dependent correction we observe that $\l_1$ is stable and practically not affected by the introduction of $\l_3$. On the other hand the central value of $\l_2$ manifests a sensible change, which is evidently due to the fact that both $\l_2$ and $\l_3$ cooperate to improve the description of the low-energy data.
Looking at Tab.~\ref{tab:tevlow_mstw08_q0pt_pointsQ}  and Tab.~\ref{tab:CTEQ3} one sees also that the Tevatron data ($\chi^2/{\rm points}\ll 1$) are probably over-parametrized in a NNLL fit with $\l_3\neq 0$. This is due to the fact that the non-perturbative effects for $Z$-boson production are probably important only for the small fraction of  data at low transverse momentum. Also this fact is not evident when looking just at the NLL fit. Notice that the range of $q_T$ explored in this work is basically the same as the one considered by different authors (see, for instance, Ref.~\cite{Landry:2002ix}).

\begin{figure}[h]
 \begin{center}
\includegraphics[width=0.5\textwidth, angle=0,natwidth=610,natheight=642]{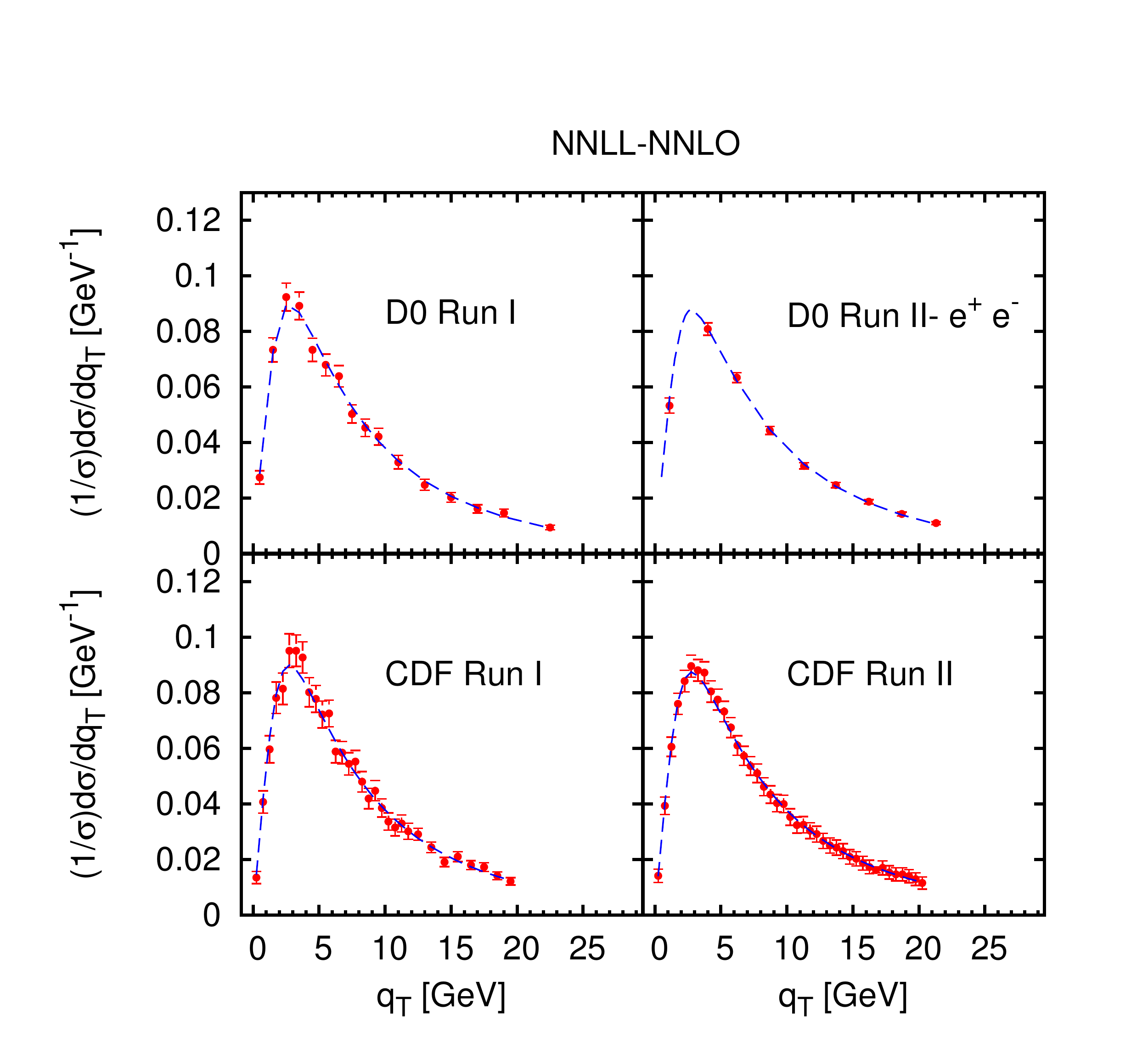} \vspace*{-.1cm}\includegraphics[width=0.4\textwidth, angle=0,natwidth=610,natheight=642]{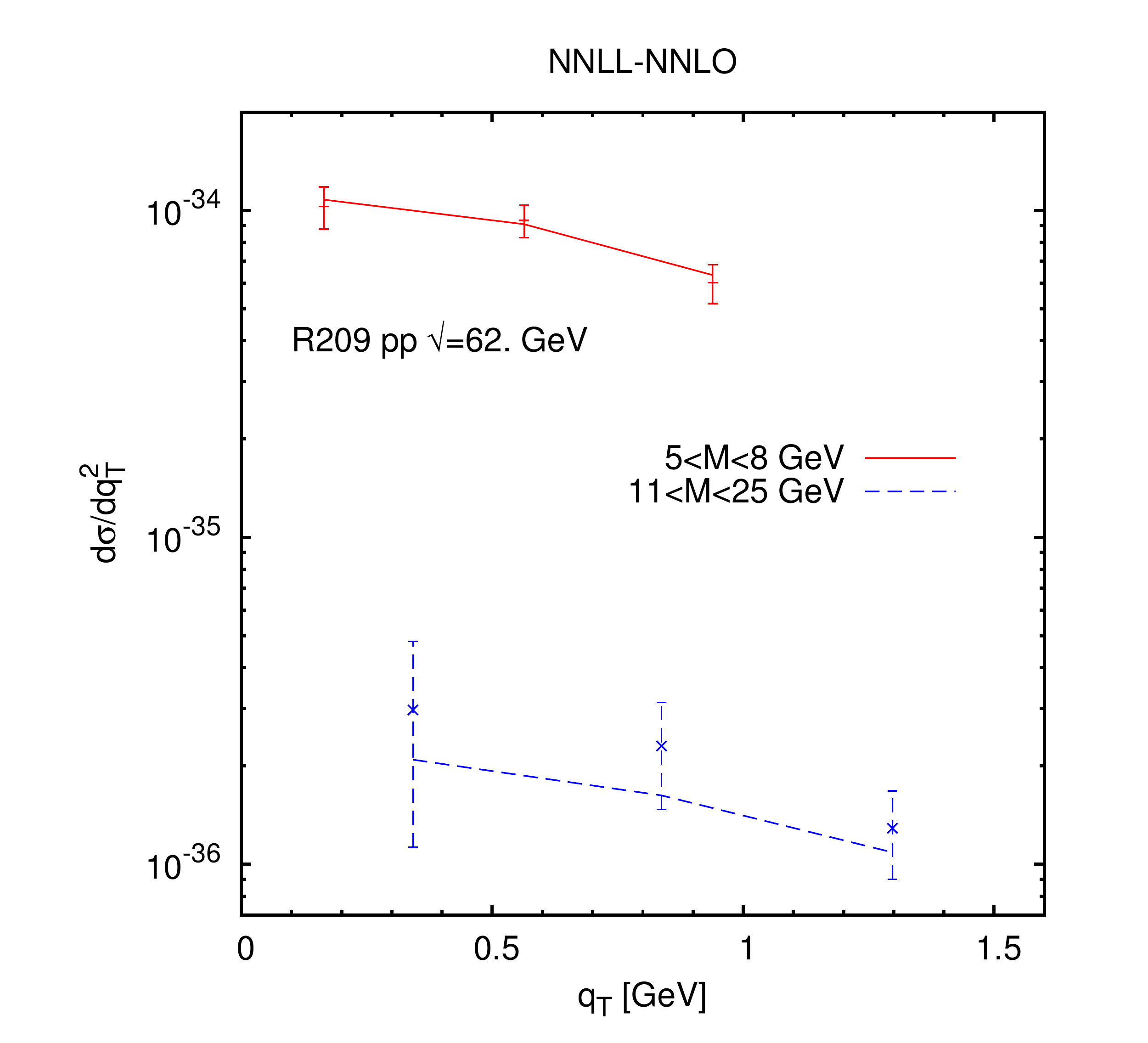}\\
\vspace*{.2cm}
 \includegraphics[width=.8\textwidth, angle=0,natwidth=610,natheight=642]{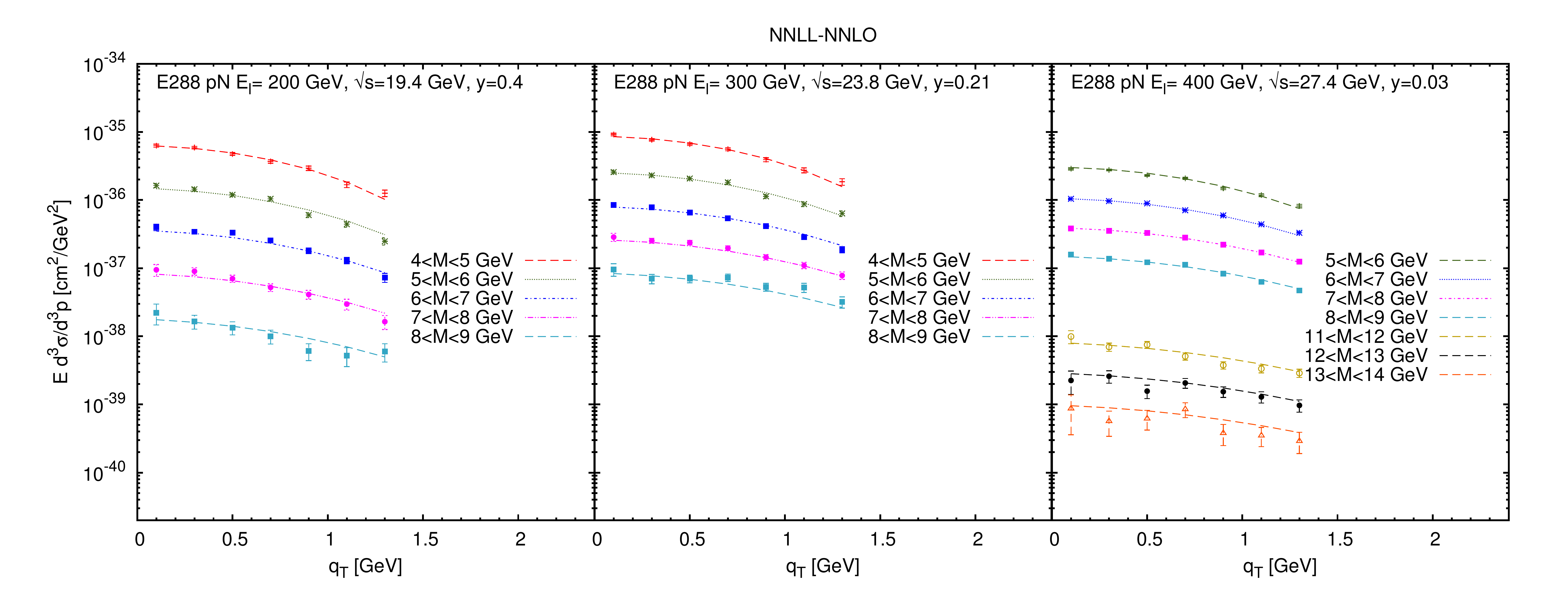}
 \caption{Best-fit curves for the analysis with $D^{\rm NP}\ne0$ (Eq.~(\ref{eq:FqN3m})), at NNLL accuracy. Comparison with Tevatron data (upper-left panel), with R209 data (upper-right panel) and E288 data (lower panels). See text for details.
\label{fig:l3}}
 \end{center}
 \end{figure}

\begin{figure}
\begin{center}
\includegraphics[width=0.6\textwidth, angle=0,natwidth=610,natheight=642]{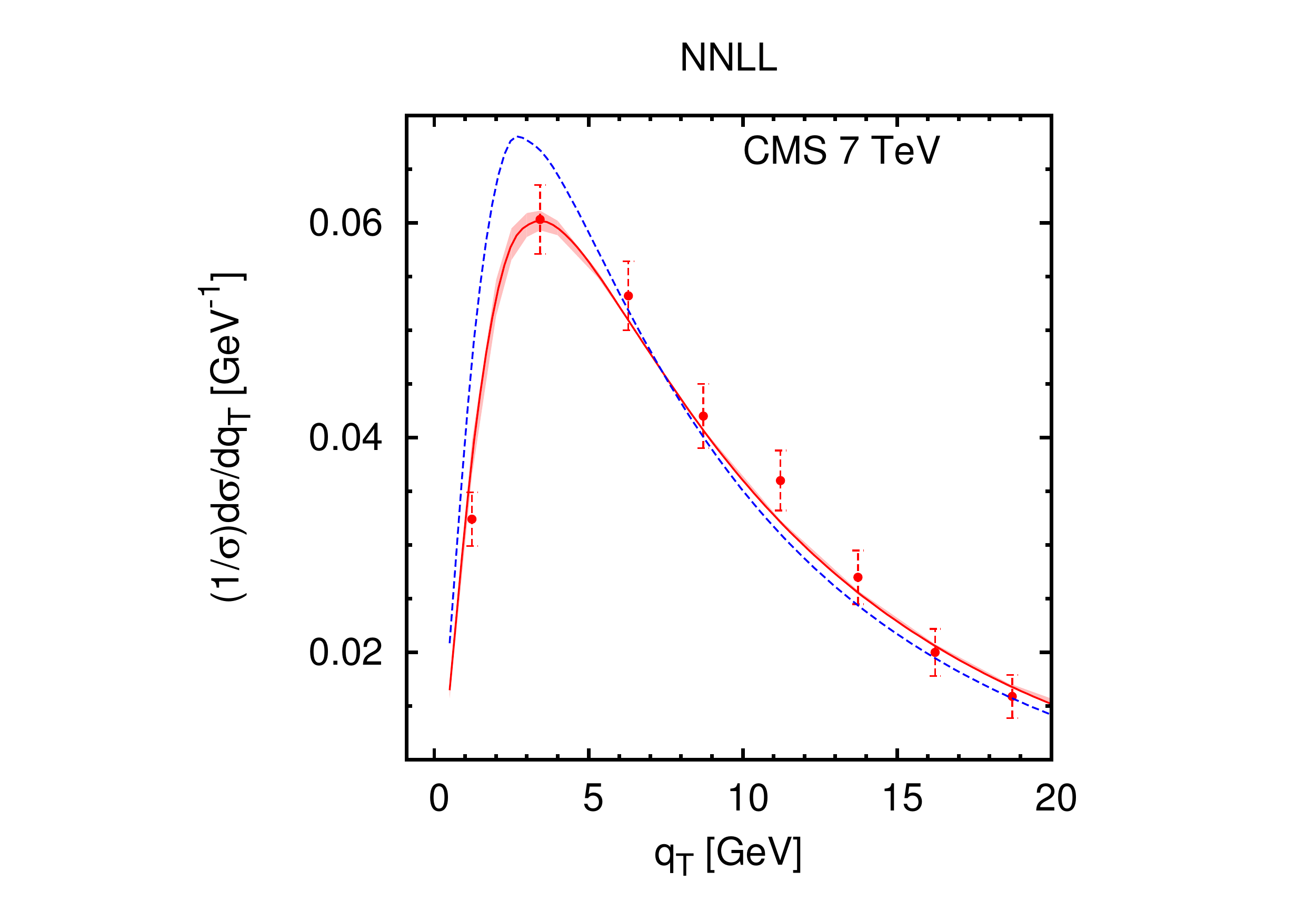
}
\caption{Our prediction (red solid line) for $(1/\sigma)d\sigma/dq_T$ based on the global fit with $D^{\rm NP}=0$ (Eq.~(\ref{eq:FqN3})), \mbox{$Q_i=Q_0+q_T$}, at NNLL-NNLO accuracy compared to CMS experimental data~\cite{Chatrchyan:2011wt}. The band comes from the statistical error on the fitted parameter $\l_1$.
The blue dashed line is the full resummed result at NNLL-NNLO accuracy with no non-perturbative input, $\la_1=\la_2=0$. For the collinear PDFs we use the MSTW08 set~\cite{Martin:2009iq}.}
\label{fig:cms7tev_pred_nnll_nnlo_mstw08_q0pt}
\end{center}
\end{figure}

For completeness in Fig.~\ref{fig:l3} we show the best-fit curves at NNLL for this model.
In the case of Tevatron data we find that the peak region is slightly enhanced providing so a better description of the data.

We conclude by pointing out that the parameter $\l_3$ parametrizes a non-perturbative correction to the evolution kernel and as such, it is spin independent and the same for TMDPDFs and TMDFFs.
Thus we expect that this correction should be included also in the analysis of SIDIS data.
Further study in this direction is left for future work.

\subsubsection*{Predictions}

As a first application and test of this phenomenological analysis, we use the present TMD formalism and our estimates at NNLL-NNLO accuracy to predict the $q_T$ dependence for $Z$-boson production at CMS (here we use the model with $\l_3=0$). Our results, shown in Fig.~\ref{fig:cms7tev_pred_nnll_nnlo_mstw08_q0pt}, are compared with the data from Ref.~\cite{Chatrchyan:2011wt}. In the plot the red solid curve is our prediction based on the fully resummed cross section at NNLL-NNLO accuracy and including the non-perturbative model.
The red band corresponds to our error on the non-perturbative input (more precisely to the statistical uncertainty on $\l_1$), being the scale error on the fitted parameters much smaller.
On the same plot the blue dashed line describes the pure perturbative fully resummed cross section without the non-perturbative model, i.e. with $\l_{1,2}=0$.
The outcome of this result is that the non-perturbative inputs are necessary to describe the peak region, $q_T \lesssim 5-10$ GeV, while the fully resummed result would be sufficient for higher values of the transverse momentum. See next Section for a more detailed discussion on these issues.


\subsection{Stability of the final results}
\label{sec:scakes}

In the previous Sections we have discussed, in some detail, the role of the scale dependence as well as of the theoretical errors on the fitted (non-perturbative) parameters. Here, to corroborate the reliability of our findings, we study the impact of the theoretical uncertainties on the stability of our results. Notice that due to the still inconclusive role of the $Q$-dependent non-perturbative factor $D^{\rm NP}$ in the fit, here below we adopt the expression of the non-perturbative model as given in Eq.~(\ref{eq:FqN3}), i.e. with $\l_3=0$. Similar results are also valid for the model with $D^{\rm NP}\ne 0$ (Eq.~(\ref{eq:FqN3m})).

We focus then on two relevant key issues, to be considered consistently to assess the validity of this type of analysis: $i)$ a comparison between the NLL and the NNLL fits including the non-perturbative model; $ii)$ a comparison of our full NNLL results with those given by pure perturbative calculations, $\la_1=\la_2=0$ (at the same order of accuracy). The aim of this study is, firstly, to check the convergence of the series (with its proper resummation) and, secondly, after showing to which extent the first aspect is understood, address the relevance of the non-perturbative inputs in the understanding of current data. In other words, the extraction of the non-perturbative pieces is sound only if the stability of the perturbative calculations is under control.

\begin{figure}
\begin{center}
\includegraphics[width=0.6\textwidth, angle=0,natwidth=610,natheight=642]{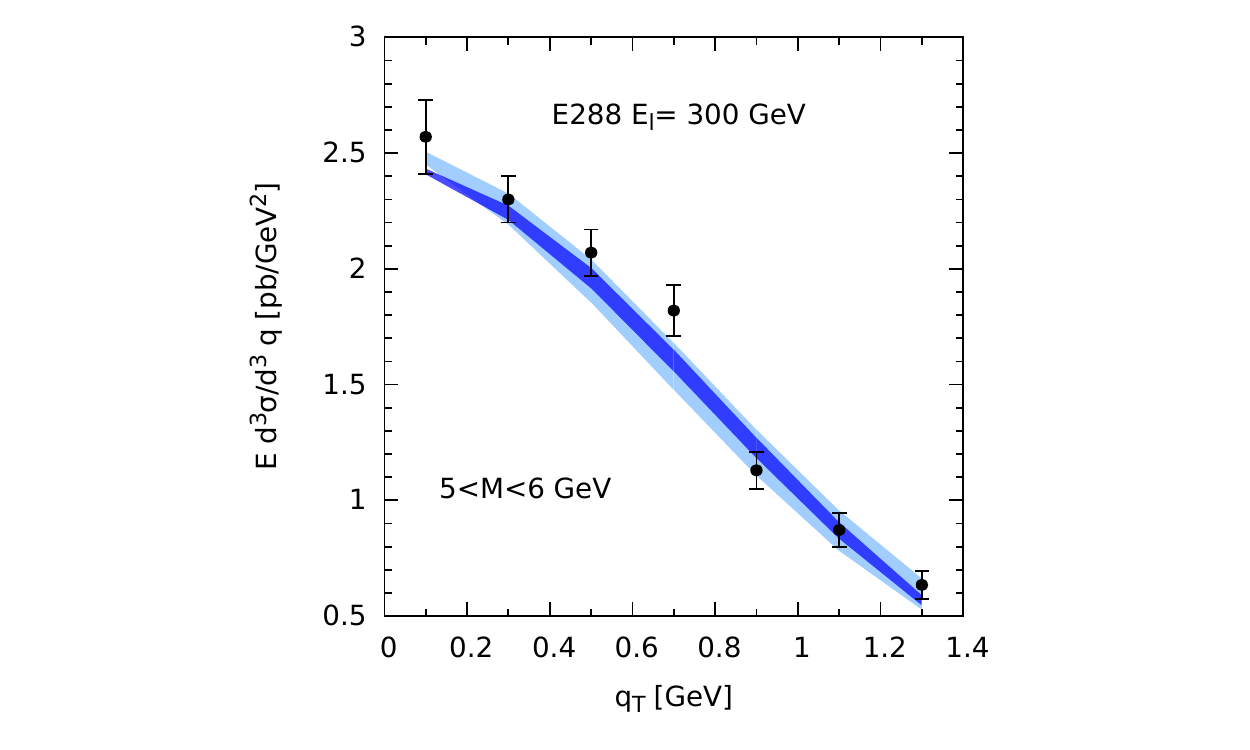}
\hspace{-4cm}
\includegraphics[width=0.6\textwidth, angle=0,natwidth=610,natheight=642]{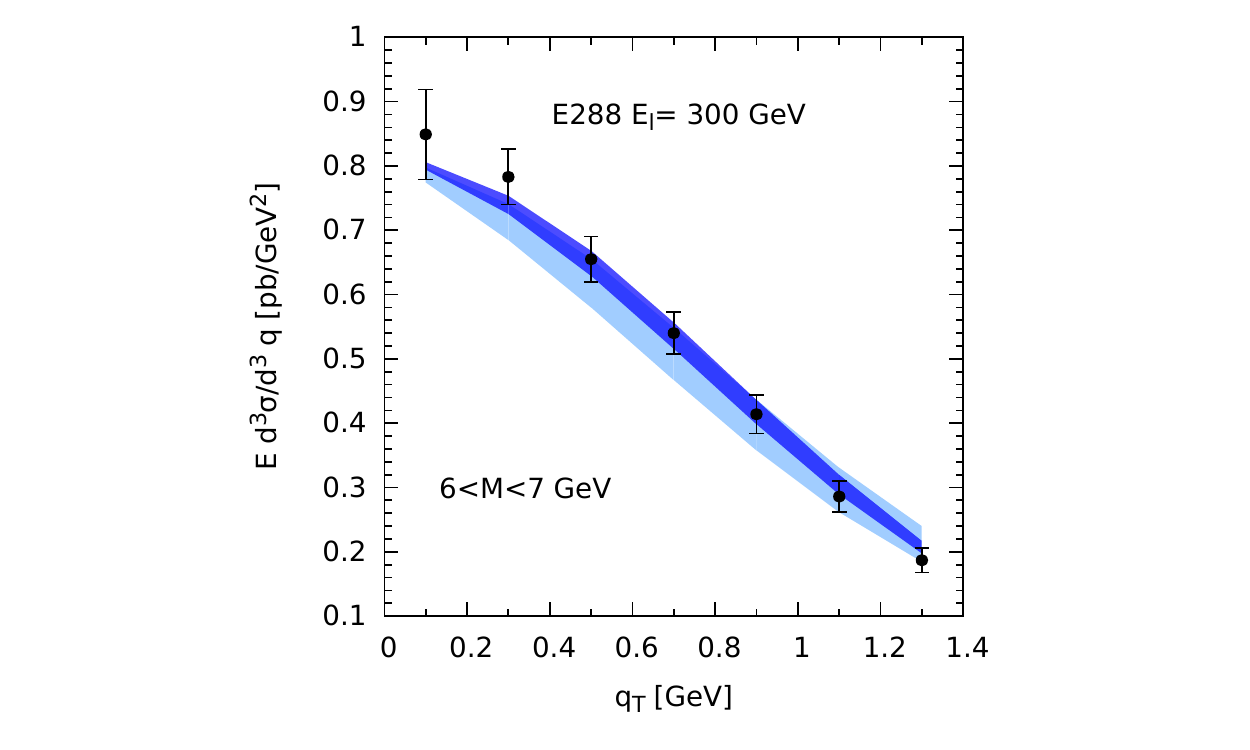}
\hspace{-4cm}
\includegraphics[width=0.6\textwidth, angle=0,natwidth=610,natheight=642]{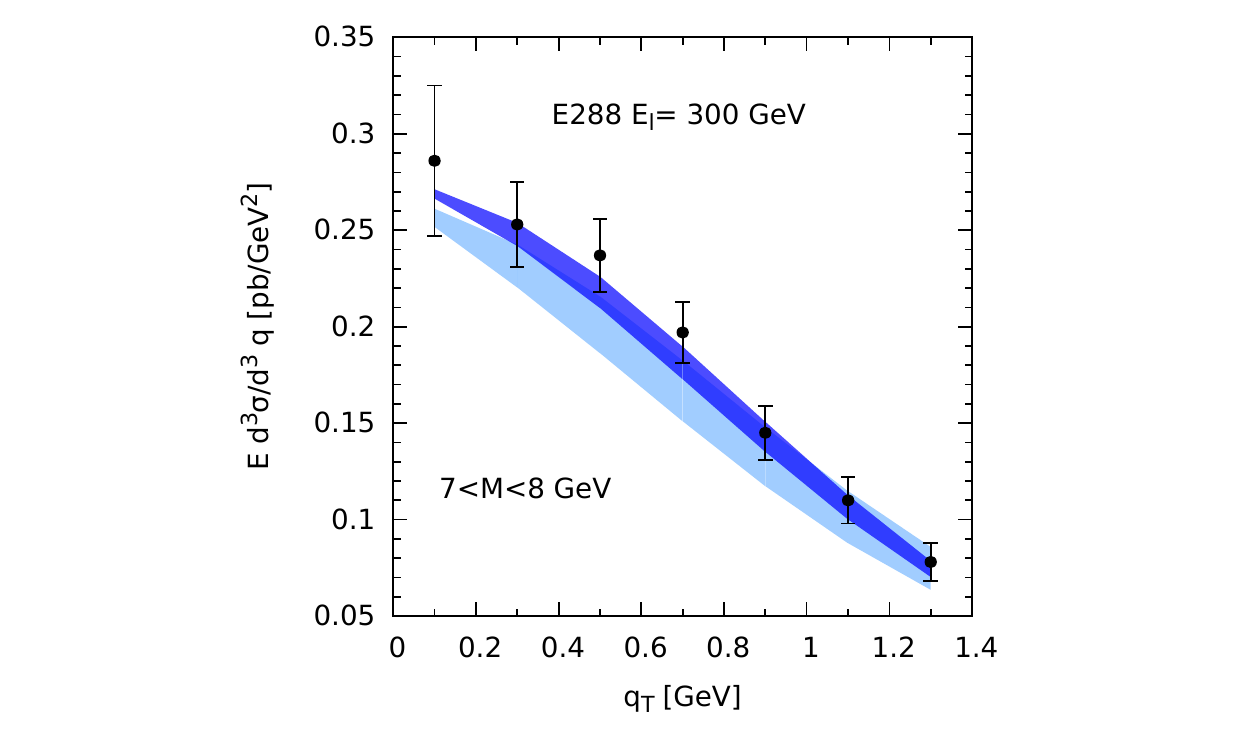}
\hspace{-4cm}
\includegraphics[width=0.6\textwidth, angle=0,natwidth=610,natheight=642]{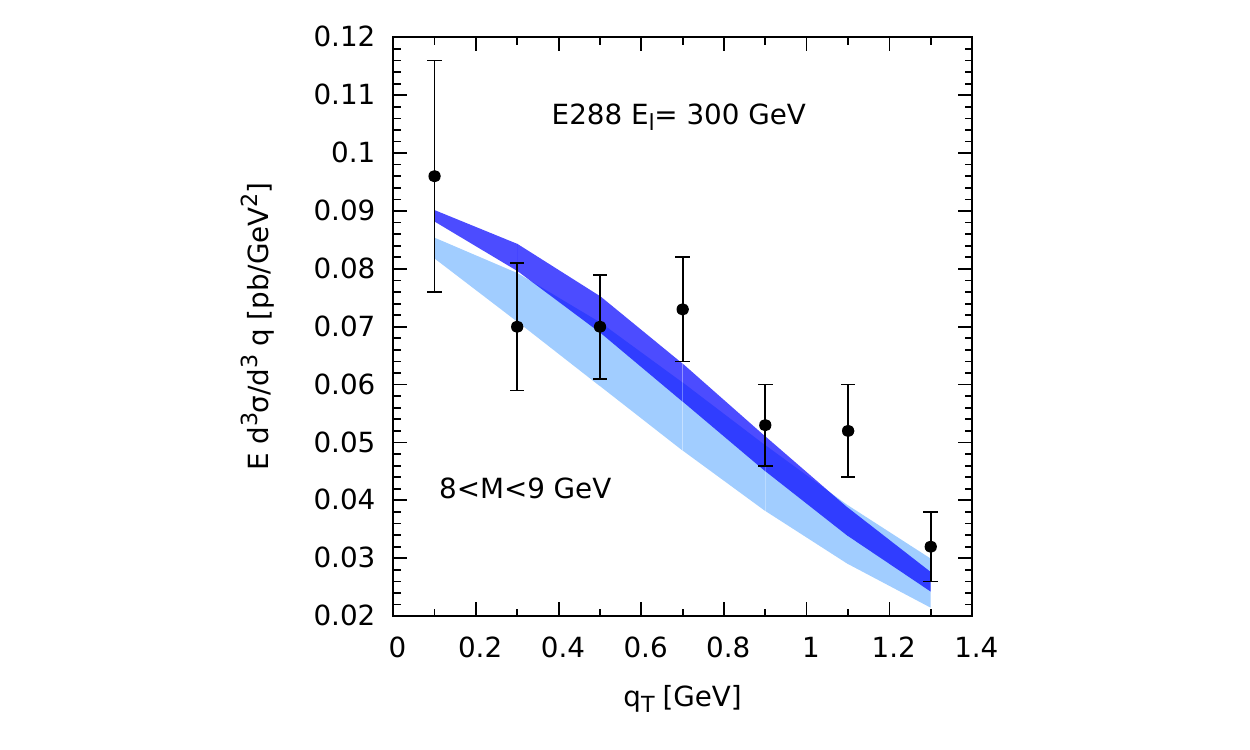}
\hspace{-4cm}
\includegraphics[width=0.6\textwidth,
angle=0,natwidth=610,natheight=642]{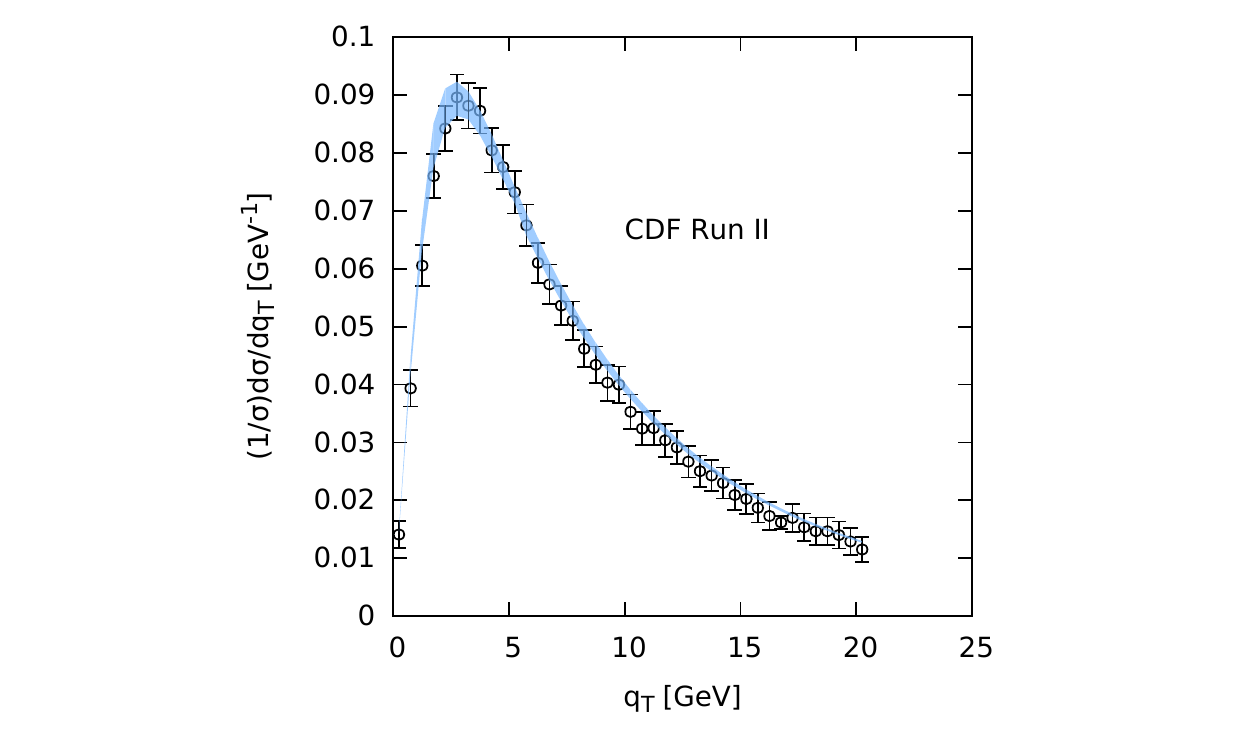}
\hspace{-4cm}
\includegraphics[width=0.6\textwidth,
angle=0,natwidth=610,natheight=642]{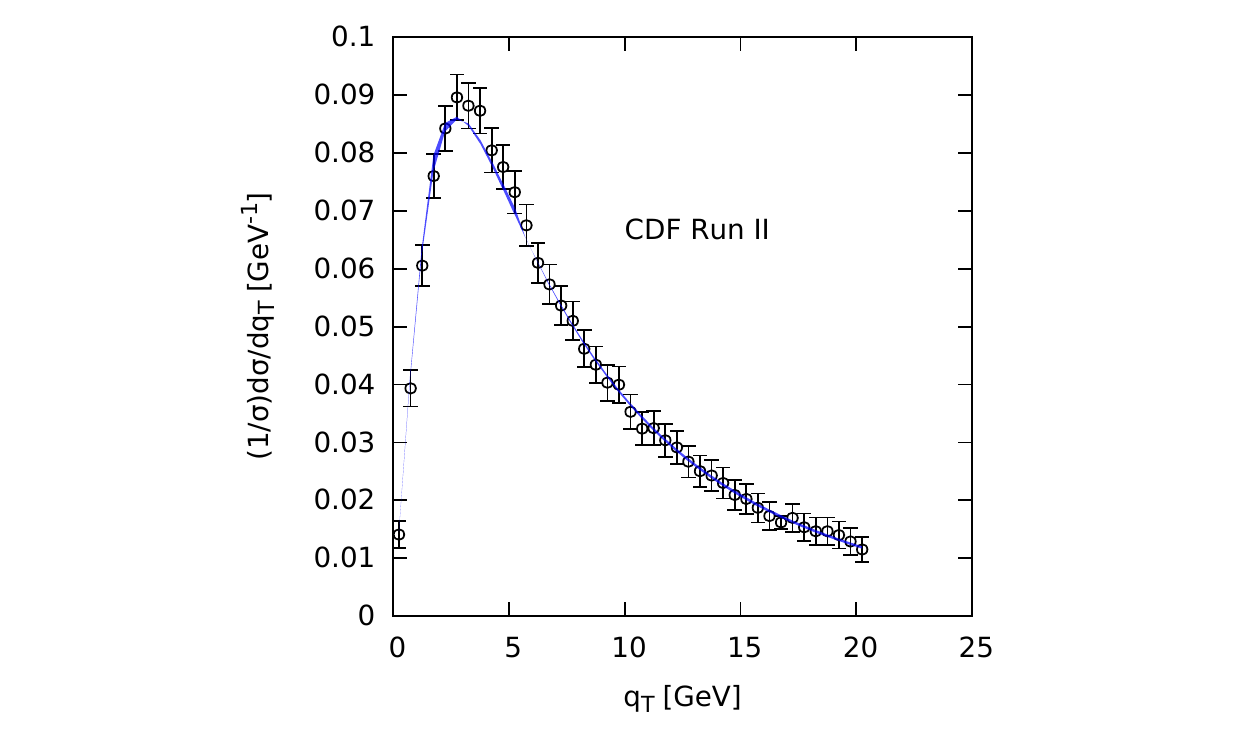}
\caption{Four upper panels: scale dependence of the NLL (cyan bands) and NNLL (blue bands) fits for four E288 data sets at $E_l= 300$ GeV. Lower panels: same study but for $Z$-boson production at CDF-Run II of NLL (left) and NNLL (right) fits. The model with $D^{\rm NP}= 0$ (Eq.~(\ref{eq:FqN3})) is used. For the collinear PDFs we use the MSTW08 set~\cite{Martin:2009iq}.}
\label{fig:NNLLvsNLL_SCALE_ERR}
\end{center}
\end{figure}

To address the first issue we compare the scale dependence at NNLL and NLL accuracies, keeping $Q_0 = 2$ GeV and varying $Q_0 + q_T /2 < Q_i< {\rm min} (Q_0 + 2q_T ;Q)$, both for low-energy DY and $Z$-boson production: the results are shown in Fig.~\ref{fig:NNLLvsNLL_SCALE_ERR}.
Here the {\em wider} cyan bands correspond to the scale dependence of the NLL result, while the {\em narrower} blue bands to the NNLL calculation.
For what concerns both the DY and the $Z$-boson production cases we show the impact of the scale dependence just for a single data set, respectively, E288-300 and CDF-RUN II, as the patterns for all other experiments/sets are similar.
In all cases one can see that we achieve a reduction of the scale dependence when passing from the NLL to the NNLL approximation and that the bands at NNLL are almost completely enclosed in the NLL ones.
Regarding the $Z$-boson production the bands that we find are also consistent to the ones reported in Ref.~\cite{Becher:2011xn}. Notice that for the Tevatron $Z$-production fits both the NLL and NNLL curves lie within the experimental error bars for all values of the transverse momentum.

This analysis, showing the improvement in the stability of the perturbative calculation at NNLL, leads us to the second issue: the fixing of the non-perturbative parts.

\begin{figure}
\begin{center}
\includegraphics[width=0.6\textwidth, angle=0,natwidth=610,natheight=642]
{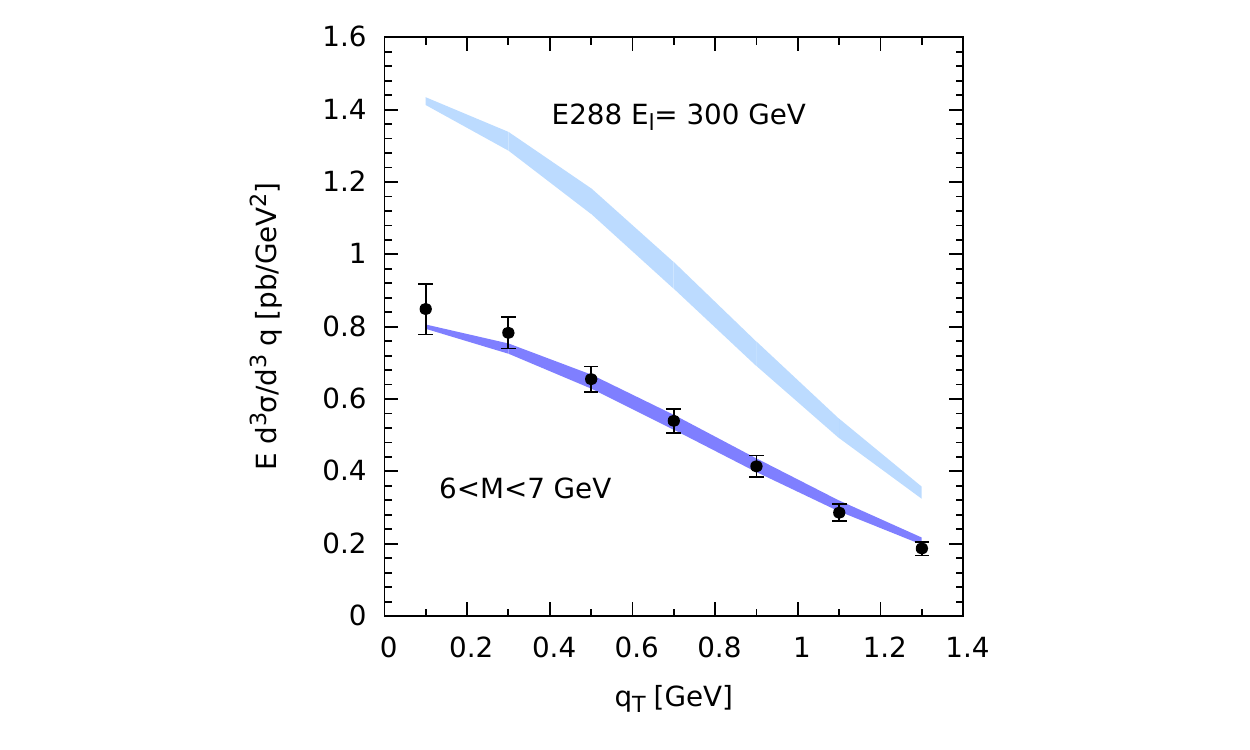}
\hspace{-4cm}
\includegraphics[width=0.6\textwidth, angle=0,natwidth=610,natheight=642]
{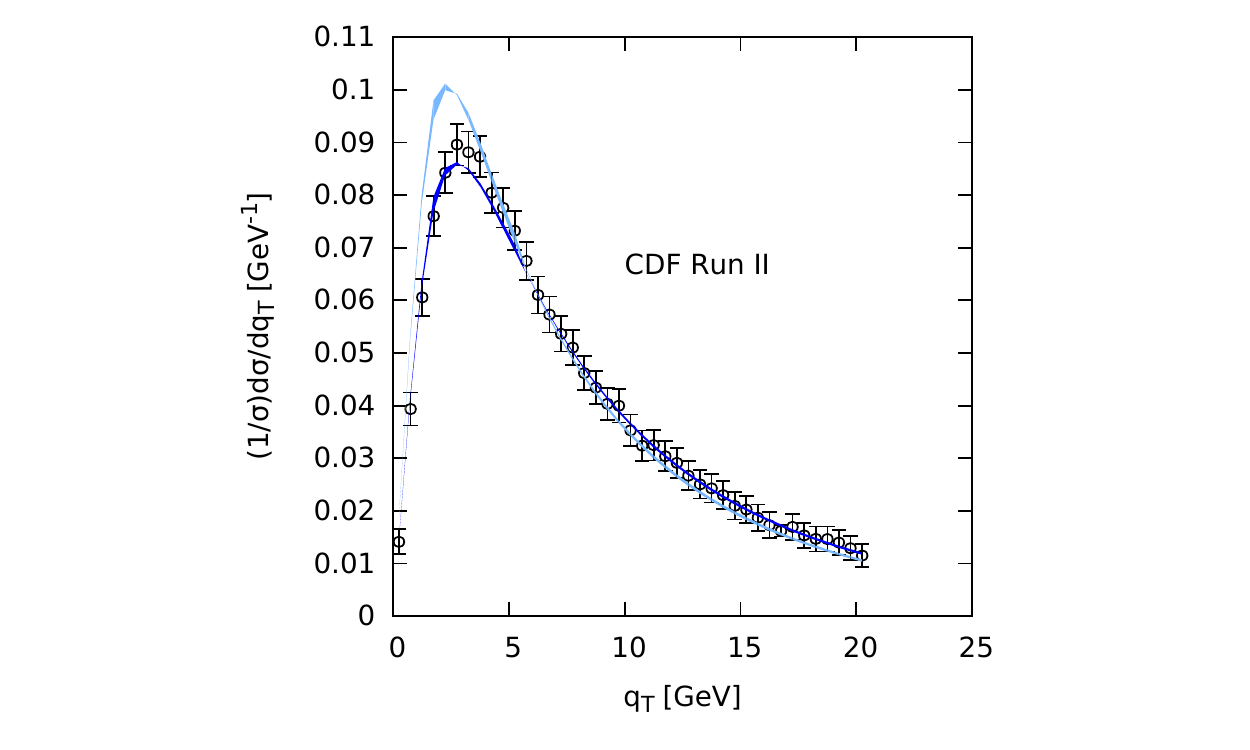}
\caption{Scale dependence of the NNLL fits (blue bands) and the pure NNLL perturbative predictions (cyan bands).
Left panel: low-energy DY at E288 with $E_l = 300$ GeV for dilepton invariant mass bin $6\;{\rm GeV}<M<7\;{\rm GeV}$. Right panel: $Z$-boson production at CDF-Run II (here the lower band corresponds to the NNLL fit). For the collinear PDFs we use the MSTW08 set~\cite{Martin:2009iq}.}
\label{fig:NPvsP_SCALE_ERR}
\end{center}
\end{figure}

To this aim we compute (and compare) our estimates {\em with} and {\em without} the introduction of a non-perturbative model at NNLL.
The result on the fitted data set is shown in Fig.~\ref{fig:NPvsP_SCALE_ERR} where the bands represent once again the effect from the scale variation.
As a first remark one can see that for low-energy DY experiments (we consider E288 as a prototype case) the pure perturbative predictions fails completely in describing the data. This can be somehow expected since this kinematic region is more sensitive to low-scale physics.
On the contrary for the $Z$-boson production case the non-perturbative effects are significative only around the peak region, as shown in Fig.~\ref{fig:NPvsP_SCALE_ERR} for the CDF-Run II experiment.

\begin{figure}
\begin{center}
\hspace*{-1.5cm}
\includegraphics[width=0.65\textwidth, angle=0,natwidth=610,natheight=642]
{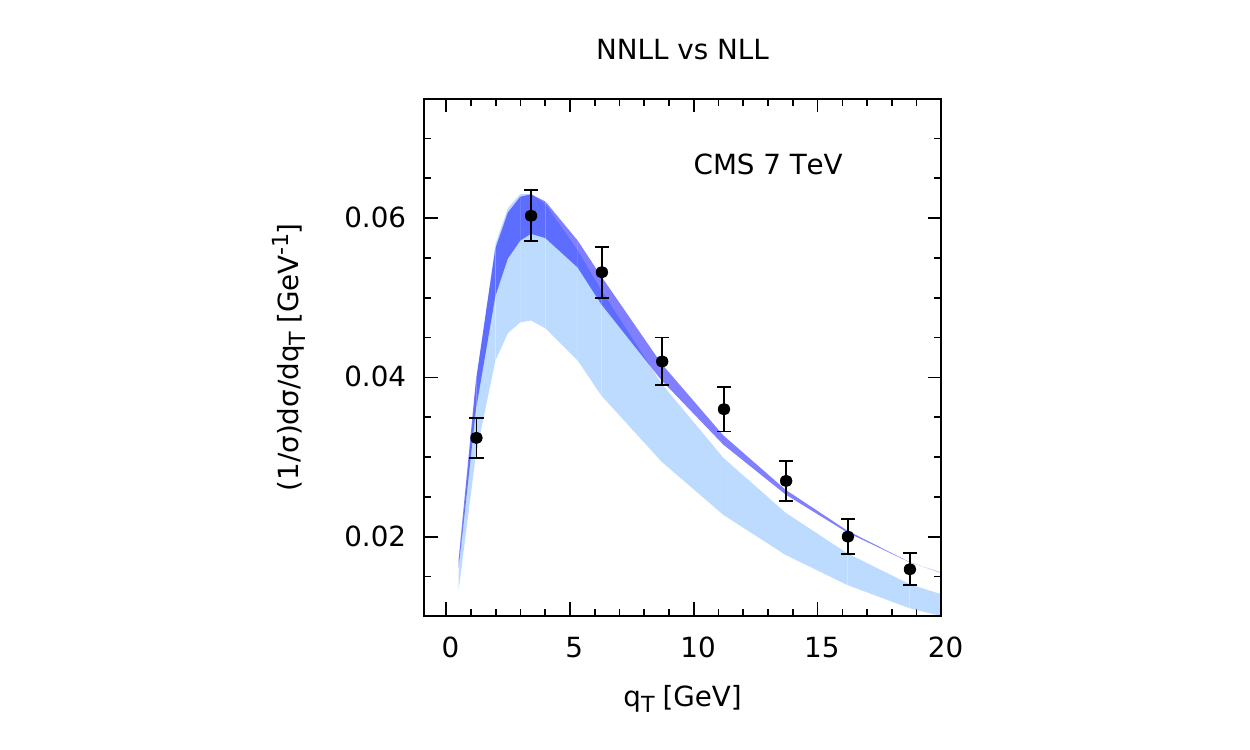}
\hspace{-5cm}
\includegraphics[width=0.65\textwidth, angle=0,natwidth=610,natheight=642]
{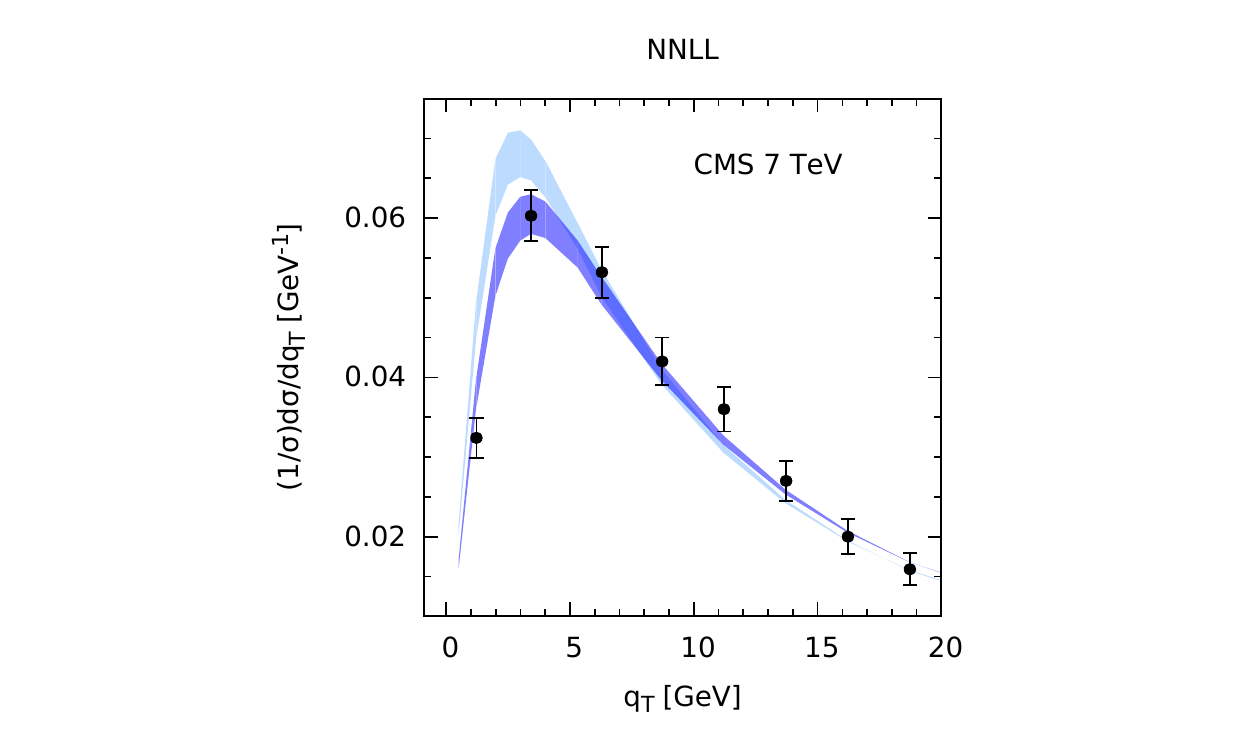}
\caption{Left panel: Impact of the scale dependence on our predictions for CMS, including the non-perturbative part of the TMDs, at NNLL (blue band) and NLL (cyan band). Right panel: Scale dependence of our predictions for CMS at NNLL with the non-perturbative part of the TMDs (blue band) and of the pure perturbative calculation with resummations (cyan band). Data are from Ref.~\cite{Chatrchyan:2011wt}. For the collinear PDFs we use the MSTW08 set~\cite{Martin:2009iq}.
  }
\label{fig:CMS_SCALE_ERR}
\end{center}
\end{figure}

Finally, to further exploit such analysis, we perform the same double-step check on our predictions for the CMS experiment~\cite{Chatrchyan:2011wt}. Notice that since CMS runs at a much higher center of mass energy with respect to the experiments considered in our global fit, different (actually smaller) values of the Bjorken variables are explored. Here, we keep using the MSTW08 PDF set~\cite{Martin:2009iq}. Similar results can be obtained with the CTEQ10 set~\cite{Lai:2010vv}.
In Fig.~\ref{fig:CMS_SCALE_ERR} (left panel) we show the impact of the scale dependence on our predictions at NLL (cyan band) and at NNLL (blue band), including the non perturbative part of the TMDs. In this case not only the NNLL achieves a reduced scale dependence with respect to NLL, but also the agreement to data is greatly improved. In Fig.~\ref{fig:CMS_SCALE_ERR} (right panel) we consider the prediction at NNLL {\em with} (blue band) and {\em without} (cyan band) the non-perturbative part of the TMDs. The error bands still come from the scale variation.
As expected the non-perturbative part of the TMDs are absolutely necessary to explain the peak region of the boson production, while its effect is diluted in the high-$q_T$ region.

From this dedicated analysis on the stability of our calculation, one can learn the following twofold message: to keep under control the perturbative part one has to consider, at least, a NNLL calculation, being the NLL approximation strongly dependent on the scale variation; on top of that, a non-perturbative piece is still necessary to reach a satisfactory description of current data and therefore it can be reliably extracted.

\section{Comparison with previous works}
\label{sec:lit}

One of the goals of this paper is to provide a framework for the analysis of transverse momentum distributions in which all ingredients coming from perturbation theory are under control and used to their maximum extent.
Only in such a case the parametrization of the non-perturbative inputs can be reliably treated.
Although such an attempt is not included in most of the studies available in the literature, different intermediate steps have been discussed in several works.
In the following we focus on the most relevant ones from our point of view.

A detailed phenomenological analysis of low- and moderate-energy DY data, aimed at extracting the transverse momentum dependence of TMDs at leading-order accuracy, appeared in Ref.~\cite{D'Alesio:2004up}. The relevance of this study was the attempt to describe within a unified, even if simplified, picture the role of TMDPDFs and TMDFFs in different hadronic processes. \par

In Ref.~\cite{Bozzi:2010xn} the authors present an analysis of transverse momentum distributions of vector boson production at NNLL.
The needed Fourier transforms to momentum space are done deforming the integration contour in $b$-space and calculating moments. They consider only very high dilepton invariant mass and explicitly avoid a complete treatment of non-perturbative effects. In this respect, within the TMD formalism that we have developed, one can achieve a direct comparison of the non-perturbative inputs with low-energy data. The evolution of TMDs allows in this way a complete fixing of this non-perturbative part, and so of the precision of the theoretical prediction.

The authors of Ref.~\cite{Becher:2011xn} perform also an analysis of $Z$-boson production. Although in their formalism they do not consider the theory for TMDs, the expression for the cross section agrees with ours when looking at the DY case or $Z$-boson production at high transverse momentum (of course they do not claim any universal structure which could eventually be used in SIDIS).
The resummations provided in our work are different from the ones in Ref.~\cite{Becher:2011xn} in the sense that in their ``collinear anomaly'' part they perform a sum of logarithms which is valid up to values of the strong coupling and impact parameter such that $\a_s L_T^2\sim 1$. Notice that one can re-obtain the ``collinear anomaly''  factor re-expanding $D^R$, $h_{\G,\g}^R$  in $\a_s$ and counting  $\a_s L_T^2\sim 1$.
Given that this is not the highest possible resummation that one can perform, the Landau pole does not appear explicitly in their resummed expression (although the perturbative series has intrinsically a Landau pole problem).
The authors in any case realize that some non-perturbative input is necessary and they suggest some Gaussian non-perturbative ($Q$-independent) part in impact parameter space, without performing any fit of $Z$-boson production data.
A non-perturbative correction to the ``collinear anomaly'' factor is suggested in~\cite{Becher:2013iya}.

In the present work we perform a complete resummation of the logs with the counting $\a_s L_T\sim 1$, which is more relevant when low-energy data are included, and we check that an exponential non-perturbative ($Q$-independent)  correction in impact parameter space works better than the Gaussian one for the $Z$-production data.
A more thorough analysis is then done here to describe also the low-energy DY data.
We nevertheless agree with the authors of Ref.~\cite{Becher:2011xn} that non-perturbative parameters cannot be fixed looking just at the $Z$-boson production case and that resummations allow for a better control of the scale dependence of the final result.

The most comprehensive work aimed at considering both low-energy and high-energy data and including an explicit non-perturbative parametrization of the cross section is the one of Refs.~\cite{Landry:2002ix} and~\cite{Konychev:2005iy} (the so called BLNY model) which implements a model within the standard CSS approach. Their analysis considers the cusp anomalous dimension at order $\a_s^2$ and the rest of the coefficients at order $\a_s^1$, and was done before the recent developments of the TMD formalism appeared.
They have no partial exponentiation of the matching coefficient between the TMDPDF and the PDF.
In their approach they use the well known $b^*$ prescription, where $b^*=b_T/\sqrt{1+b^2_T/b_{\rm max}^2}$ and $b_{\rm max}$ is a parameter,  modeling the $Q$-dependent non-perturbative corrections  in terms of   the parameter $g_2$, and  the $Q$-independent corrections in terms of $g_1$, $g_3$.
Consequently, as shown in Ref.~\cite{Konychev:2005iy}, all the parameters $g_{1,2,3}$ heavily depend on the choice of $b_{\rm max}$. In that work they find as a best value  $b_{\rm max}\sim 1.5$ GeV$^{-1}$, consistent with the low-$b_T$ behavior of $D^R$ as observed in Ref.~\cite{Echevarria:2012pw}. Notice that the parameter $g_2$ is the same for DY, SIDIS and $e^+e^-$ processes and for all leading-twist TMDs, while $g_{1,3}$ are specific of the unpolarized TMDPDF.
An important point is that the DY data from which these parameters are extracted have $Q\geq 4$ GeV, while SIDIS data (say from HERMES or COMPASS experiments) on which these parameters are often used, cover lower $Q^2$ ranges.
Notice that in BLNY the model of the evolution kernel (which is $Q$-dependent) plays a fundamental role and is crucial in the fit of data.

The point of view of our work is instead completely different.
In fact, we have shown that in order to analyze the Drell-Yan data (i.e.~for dilepton invariant mass $Q\geq 4$ GeV) the $Q$-dependent non-perturbative corrections play a minor role and most of the non-perturbative effects are $Q$-independent.
In other words, we have found that the extraction of the parameters $b_{\rm max}$ and $g_2$ in the BLNY model is done using data for which the evolution kernel can instead be almost entirely predicted theoretically.
The differences in our fit between the cases with $\l_3=0$ and $\l_ 3\neq 0$ are important but not essential for the success of the fit.
In any case, the final $\chi^2$ that we get with $\l_3\neq 0$ improves notably the results of BLNY.
A second issue concerns the fact that the parameters $(b_{\rm max},\; g_2)$ extracted from DY data are used at scales   different from those of their extraction.
It is possible that using these values in describing low energy SIDIS data (which usually have $1$ GeV $<Q<4$ GeV) can cause some tension.
While this problem could be avoided with our parametrization of the TMDs, it needs a dedicated study which is beyond the scope of this paper and we leave for future work.
Finally for a closer comparison to the fit of Ref.~\cite{Landry:2002ix}, where they include also the two bins of E605 with the lowest dilepton mass, we have also checked the impact of these two bins on our global fit with $\l_3\neq 0$ at NNLL. The number of points of the two bins is 14, and so the impact of these points on our global fit is relatively small. The central values  and the errors of $\l_{1,2,3}$ do not change while the  final $\chi^2/{\rm d.o.f.}=0.95$, confirming  our considerations in Section~\ref{data sel}.

Several modifications of the BLNY model have been proposed in the literature and sometimes used in actual fits.
In Ref.~\cite{Aidala:2014hva} the authors propose a different parametrization for the $g_2$ term, based on phenomenological considerations of SIDIS data. Again, we can apply the same considerations as in the original BLNY model: while we do not exclude a non-perturbative structure of the evolution kernel, its relevance cannot be clearly stated doing a NLL fit given the current status of DY data.

In Ref.~\cite{Echevarria:2014xaa} a first attempt of a global fit of SIDIS and DY data is performed, by using the $b^*$ prescription and resumming large logarithms at NLL accuracy.
The fit of data is only qualitatively correct and the final $\chi^2$ is not declared.
In the present work we have shown the importance of including higher perturbative orders and resummations in the parametrization of the TMDPDFs, together with other ingredients.

Recently a fit of DY data has been performed in Ref.~\cite{Su:2014wpa}.
The authors propose a modification of the BLNY model, changing both the $Q$-dependent part (i.e.~the evolution kernel), and the $Q$-independent part (inserting new arbitrary parameters).
Concerning the fit of the DY data they fit the three standard parameters of the BLNY model plus the normalization of all experimental data sets they use.
The fit is done considering  only a subset of low energy data.
Moreover the partial $\chi^2$'s of each experiment are not shown and the total $\chi^2$ for the DY data declared is about 1.5.
When they consider SIDIS data they increase the number of parameters, namely the ones for the TMD fragmentation function, and a normalization factor for each $z$-bin.
Given the amount of variables in the fit, the almost arbitrary choice of data and the final $\chi^2$ they get, their result cannot be considered conclusive.

\section{Conclusions}
\label{concl}

The TMD formalism is a powerful tool to analyze perturbative and non-perturbative effects in $q_T$ spectra.
In this work we have fitted the DY and $Z$-boson production data to fix the non-perturbative part of TMDPDFs: all results are collected in Tab.~\ref{tab:tevlow_mstw08_q0pt_param},  \ref{tab:tevlow_mstw08_q0pt_paramQ} and Tab.~\ref{tab:CTEQ2}, \ref{tab:CTEQ3}.
We have stressed the fact that in order to have a reliable fixing of the non-perturbative inputs one has to provide a fully resummed expression for the perturbative part. The fully resummed cross section in fact is less sensitive to the factorization scale dependence and this allows a more stable extraction of the non-perturbative pieces of the TMDs.
We have provided a full study of the perturbative inputs. In particular we have used the TMD evolution kernel at NNLL~\cite{Echevarria:2012pw} which, to our knowledge, was never used before in a global fit of this kind.
We have also discussed the matching of TMDPDFs onto PDFs with the exponentiation and fully resummation of the  corresponding coefficient. We have argued that the exponentiated part of this matching coefficient is spin independent and should be included in the analysis of other types of TMDPDFs. This part is fundamental to have a reliable description of the TMDs both at NLL and NNLL.

One of the important aspects of our perturbative analysis is that the factorization scale is fixed in momentum space  instead of the more usual impact parameter space.
This choice provides a good stability of the perturbative series and offers a new understanding of the data and of the model dependence of the TMDs.
We find that the NNLL fit clarifies several issues about the non-perturbative nature of TMDs.

The model-dependent non-perturbative inputs for the TMDPDF are studied in order to minimize the number of non-perturbative parameters and to provide a good description of the data.
We find that the $Z$-boson data are better described by an exponential damping factor in impact parameter space rather than a Gaussian one. The associated parameter, called here $\l_1$, has a stable value within the errors, which are mainly of statistical origin.
The low-energy data explore values of the impact parameter higher than those covered in the case of $Z$-boson production.
We find that a polynomial correction with a new parameter, called $\l_2$, plays a relevant role in this respect and both corrections, induced by $\l_{1,2}$, do not depend on the dilepton invariant mass $Q$.
The values of these parameters can be fixed by fitting data for DY and $Z$-boson production and the NNLL resummation greatly reduces the theoretical error on this determination.

Particular attention has been paid to the study of the $Q$ dependence of the non-perturbative model. The insertion of this contribution (parametrized by $\l_3$) provides only an improvement of the $\chi^2$ at the price of adding a new parameter to the fit.
Nevertheless, given the actual uncertainties on data and collinear PDFs, the need for this correction in the fits cannot be firmly established.
Increasing the precision of the experimental data can be crucial to fix this issue.

One important outcome of this work is that a fully resummed evolution kernel, together with other exponentiations and resummations in the various matching coefficients that appear in the cross section, avoid an excessive use of a modelization of the cross section, making the predictions more stable.

We consider this work as a first step towards the proper understanding of non-perturbative effects in transverse momentum distributions. Several important perturbative pieces, recently calculated~\cite{Catani:2011kr,Catani:2012qa,Gehrmann:2012ze,Gehrmann:2014yya}, can be used in an approximate N$^3$LL analysis and will be included in a forthcoming publication.

We point out that fixing the non-perturbative part of transverse momentum distributions can improve substantially the theoretical precision needed for the current LHC experiments, as our prediction for $Z$-boson $q_T$ spectrum at CMS shows. In order to have a complete understanding of the result of this experiment  we need both a description of the non-perturbative part of the TMDs and a NNLL resummation, as shown by this study (see, e.g., Figs.~\ref{fig:cms7tev_pred_nnll_nnlo_mstw08_q0pt} and \ref{fig:CMS_SCALE_ERR}).

Finally, we comment on the use of this formalism for SIDIS processes.
The parameters $\l_{1,2}$ are specific of the unpolarized TMDPDF and can be used also in the analysis of SIDIS data.
A different value of these parameters is expected for the fragmentation function. The parameter $\l_3$ is a universal correction and, as such, it is the same in DY and SIDIS processes.
One important aspect that must be pointed out is that most of SIDIS data at our disposal are given for 1 GeV $< Q <$ 4 GeV, that is in an energy range lower than the one studied in this work.
For this reason, it can be that the model with $D^{\rm NP} \neq 0$ (maybe different from the one discussed in this work) is  better suited for the description of the SIDIS data in this regime and we expect  that a NNLL study would be fundamental to understand them.

\section*{Acknowledgements}
U.D.~is grateful to the Department of Theoretical Physics II of the Universidad Complutense of Madrid for the kind hospitality extended to him since the earlier stages of this work. I.S.~would like to thank the CERN where the initial part of this work has been done and early comments of G.~Altarelli, S.~Frixione, A.~Idilbi and G.~Salam.
M.G.E.~is supported by the ``Stichting voor Fundamenteel Onderzoek der Materie'' (FOM), which is financially supported by the ``Nederlandse Organisatie voor Wetenschappelijk Onderzoek'' (NWO).
U.D.~and S.M.~acknowledge support from the European Community under the FP7 program ``Capacities - Research Infrastructures'' (HadronPhysics3, Grant Agreement 283286). S.M.~is partly supported by the ``Progetto di Ricerca Ateneo/CSP'' (codice TO-Call3-2012-0103).
I.S.~is supported by the Spanish MECD grant, FPA2011-27853-CO2-02.

\appendix*
\section{Fits with CTEQ10 PDF set}
\label{app:cteq}
We collect here the corresponding results of our fits at NLL and NNLL accuracies obtained using the CTEQ10 set~\cite{Lai:2010vv} for the collinear parton distributions.

\begin{table}[h]
\begin{tabular}{|c|c|c|c|c|c|}
 \hline
      ~        &   ~        &  NNLL, NNLO       &    NLL, NLO \\
 \hline
     ~        &    points  &   $\chi^2/\textrm{points}$  &    $\chi^2/\textrm{points}$   \\
 \hline
              &    223     &      0.96          &     1.79           \\
 \hline
 \hline
 E288 200     &     35    &      1.58            &    2.61      \\
  \hline
 E288 300     &     35     &      1.09          &    1.10       \\
 \hline
 E288 400     &     49     &      1.17          &    2.43     \\
 \hline
 \hline
 R209         &      6    &       0.20           &     0.35            \\
 \hline
 \hline
  CDF Run I   &     32    &       0.83           &     1.55                \\
 \hline
  D0 Run I    &     16     &       0.48          &     1.79      \\
 \hline
  CDF Run II  &     41     &       0.38          &     0.79    \\
 \hline
  D0 Run II   &      9     &       1.036          &     3.28   \\
 \hline
\end{tabular}
\caption{
Total and partial $\chi^2/\textrm{points}$ of our global fit on low-energy~\cite{Ito:1980ev,Antreasyan:1981uv} and Tevatron data~\cite{Affolder:1999jh,Abbott:1999yd,Abbott:1999wk, Aaltonen:2012fi, Abazov:2007ac} with $D^{\rm NP}=0$ (Eq.~(\ref{eq:FqN3})), \mbox{$Q_i=Q_0+q_T$}, at NNLL and NNL accuracies and with the collinear PDFs from  CTEQ10~\cite{Lai:2010vv}, respectively at NNLO and NLO.
\label{tab:CTEQ1}
}
\end{table}

\begin{table}[h]
\renewcommand{\tabcolsep}{0.4pc} 
\renewcommand{\arraystretch}{1.2} 
\begin{center}
\begin{tabular}{|l|l|l|}
 \hline
          NLL                       &  223  points           &   $\chi^2$/dof = 1.79  \\
  \hline
 ~                                  &  $\lambda_1$ = $0.28\pm0.05_{\rm stat}\textrm{ GeV}$ &   $\lambda_2=0.14\pm0.04_{\rm stat}\textrm{ GeV}^2$  \\
\hline
~                                   &  $N_{\rm E288}=1.02\pm0.04_{\rm stat}$&$N_{\rm R209}=1.4\pm0.2_{\rm stat}$\\
\hline
\hline
          NNLL                       &  223  points           &   $\chi^2$/dof = 0.96  \\
  \hline
 ~                                  &  $\lambda_1$ = $0.32\pm0.05_{\rm stat}\textrm{ GeV}$ &   $\lambda_2=0.12\pm0.03_{\rm stat}\textrm{ GeV}^2$  \\
\hline
~                                   &  $N_{\rm E288}=0.99\pm0.05_{\rm stat}$&$N_{\rm R209}=1.6\pm0.3_{\rm stat}$\\
\hline
\end{tabular}
\caption{
Results of our global fit on low-energy~\cite{Ito:1980ev,Antreasyan:1981uv} and Tevatron data~\cite{Affolder:1999jh,Abbott:1999yd,Abbott:1999wk, Aaltonen:2012fi, Abazov:2007ac}, with $D^{\rm NP}=0$ (Eq.~(\ref{eq:FqN3})), \mbox{$Q_i=Q_0+q_T$}, at NNLL and NNL accuracies and with the collinear PDFs from   CTEQ10~\cite{Lai:2010vv}, respectively at NNLO and NLO.
\label{tab:CTEQ2}}
\end{center}
\end{table}

\begin{table}[h]
\vskip 18pt
\renewcommand{\tabcolsep}{0.4pc} 
\renewcommand{\arraystretch}{1.2} 
\begin{tabular}{|c|c|c|c|c|}
 \hline
 $Q_0=2.0 \textrm{ GeV}+q_T$ &~ &     NNLL               &           NLL           \\
 \hline
 \hline
   $\lambda_1$    &   ~   & $0.29\pm0.04_{\rm stat}\textrm{ GeV}$   & $0.27\pm0.06_{\rm stat}\textrm{ GeV}$    \\
  \hline
   $\lambda_2$    &   ~   & $0.170\pm0.003_{\rm stat}\textrm{ GeV}^2$ & $0.19\pm0.06_{\rm stat}\textrm{ GeV}^2$   \\
  \hline
    $\lambda_3$    &   ~   & $0.030\pm0.01_{\rm stat}\textrm{ GeV}^2$ & $0.02\pm0.01_{\rm stat}\textrm{ GeV}^2$   \\
  \hline
  $N_{E288}$   &    ~    &    $0.93\pm0.01_{\rm stat}$              &  $0.98\pm0.06_{\rm stat}$                              \\
 \hline
    $N_{R209}$ &    ~    &       $1.5\pm0.1_{\rm stat}$             &  $1.3\pm0.2_{\rm stat}$                                \\
  \hline
   $\chi^2$    &  ~          &       180.1               &    375.2                                         \\
  \hline
  \hline
  ~            &    points   &  $\chi^2/\textrm{points}$ &  $\chi^2/\textrm{points}$            \\
  \hline
               &    223      &      0.81                 &     1.68                              \\
 \hline
  ~            &    points   &  $\chi^2/\textrm{dof}$ &  $\chi^2/\textrm{dof}$            \\
 \hline
               &    223      &      0.83                 &     1.72                               \\
  \hline
  \hline
  E288 200     &     35     &       1.35                 &     2.28                            \\
  \hline
 E288 300     &     35     &       0.98                 &     1.22                          \\
 \hline
 E288 400     &     49     &       1.05                 &     2.33                           \\
 \hline
 \hline
 R209         &      6     &       0.27                 &     0.40                            \\
 \hline
 \hline
  CDF Run I   &     32     &       0.70                 &     1.50                              \\
 \hline
  D0 Run I    &     16     &       0.41                 &     1.77                                \\
 \hline
  CDF Run II  &     41     &       0.25                 &     0.76                            \\
 \hline
  D0 Run II   &      9     &       0.82                 &     3.2                          \\
 \hline
 \end{tabular}
 \caption{
Results of our global fit on low-energy~\cite{Ito:1980ev,Antreasyan:1981uv} and Tevatron data~\cite{Affolder:1999jh,Abbott:1999yd,Abbott:1999wk, Aaltonen:2012fi, Abazov:2007ac}, with $D^{\rm NP}\ne0$ (Eq.~(\ref{eq:FqN3m})), \mbox{$Q_i=Q_0+q_T$}, at NNLL and NNL accuracies and with the collinear PDFs from CTEQ10~\cite{Lai:2010vv}, respectively at NNLO and NLO.
\label{tab:CTEQ3}
}
\end{table}

\end{document}